\documentclass[notitlepage,prd,twocolumn,showpacs,preprintnumbers,nofootinbib,superscriptaddress]{revtex4-1}
\usepackage{amssymb,amsmath,placeins,epsfig,array}
\usepackage[usenames,dvipsnames]{color}
\usepackage[normalem]{ulem}
\usepackage[breaklinks,colorlinks,urlcolor=blue,citecolor=blue,linkcolor=blue]{hyperref}

\usepackage{selinput}

\def\gsim{\mathrel{\raise.3ex\hbox{$>$\kern-.75em\lower1ex\hbox{$\sim$}}}}

\newcommand{\ie}{{\it i.e.~}}  \newcommand{\eg}{{\it e.g.~}}

\newcommand{\beq}{\begin{equation}} \newcommand{\eeq}{\end{equation}}
\newcommand{\bea}{\begin{eqnarray}} \newcommand{\eea}{\end{eqnarray}}

\newcommand{\Eq}[1]{Eq.~(\ref{#1})}  
\newcommand{\Sec}[1]{Sec.~\ref{#1}}  
\newcommand{\Fig}[1]{Fig.~\ref{#1}}

\newcommand{\mb}{m_b}

\begin{document}

%\title{Photon superradiance as probe of the intergalactic medium}
%\title{On the Observability of Photon Superradiance}
\title{Quenching Mechanisms of Photon Superradiance}
\author{Diego Blas}\email{e-mail: diego.blas@kcl.ac.uk}
\affiliation{Theoretical Particle Physics and Cosmology Group, Department of Physics,\\
  ~~ King's College London, Strand, London WC2R 2LS, UK}
\author{Samuel J.~Witte}\email{e-mail: witte.sam@gmail.com}
\affiliation{Instituto de F\'{\i}sica Corpuscular (IFIC), CSIC-Universitat de Val\`encia, Spain}
\preprint{KCL-2020-55}

\begin{abstract}
Rapidly rotating black holes are known to develop instabilities in the presence of a sufficiently light boson, a process which becomes efficient when the boson's Compton wavelength is roughly the size of the black hole. This phenomenon, known as black hole superradiance, generates an exponentially growing boson cloud at the expense of the rotational energy of the black hole. For astrophysical black holes with $M \sim \mathcal{O}(10) \, M_\odot$, the superradiant condition is achieved for bosons with $m_b \sim  \mathcal{O}(10^{-11} ) \, {\rm eV}$; intriguingly, photons traversing the intergalactic medium (IGM) acquire an effective mass (due to their interactions with the ambient plasma) which naturally resides in this range. The implications of photon superradiance, \ie the evolution of the superradiant photon cloud and ambient plasma in the presence of scattering and particle production processes, have yet to be thoroughly investigated.  Here, we enumerate and discuss a number of different processes capable of quenching the growth of the photon cloud, including particle interactions with the ambient electrons and back-reactions on the effective mass (arising \eg from thermal effects, pair-production, ionization of of the local background, and modifications to the dispersion relation from strong electric fields). This work naturally serves as a guide in understanding how interactions may allow light exotic bosons to evade superradiant constraints.
\end{abstract}
%%%%%%%%%%%%%%%%%%%%%%%%%%%%%%%%%%%%%%%%%%%%%%%%%%%%%
\maketitle
%%%%%%%%%%%%%%%%%%%%%%%%%%%%%%%%%%%%%%%%%%%%%%%%%%%%%
%\newpage

%%%%%%%%%%%%%
\section{Introduction}
%%%%%%%%%%%%%

Rotational superradiance is the phenomenon in which energy can be extracted from a rotating absorbing surface via the scattering of bosons with energy $\omega < m \, \Omega \, $, where $m$ is the azimuthal number and $\Omega$ the angular velocity of the rotating body ~\cite{zel1971generation}.  It has long been appreciated that if one could confine such radiation around a Kerr black hole, \eg using a mirror located slightly outside the ergoregion, the superradiant instability would lead to a rapid and violent run-away process  --- a so-called `black hole bomb'~\cite{Press:1972zz,Teukolsky:1974yv,Cardoso:2004nk}. Remarkably, nature provides its own mirror in the form of particle mass -- that is to say, the existence of a non-zero mass induces bound states that can confine the radiation near the black hole and facilitate the instability~\cite{Damour:1976kh,detweiler1980klein,Cardoso:2004nk} (we refer to the interested reader to~\cite{Brito:2015oca} for an extensive review on rotational superradiance and its relation to black hole physics).  

The superradiant instability is maximally efficient when the gravitational radius of a black hole is approximately of the same size as the Compton wavelength of the incident boson~\cite{Witek:2012tr}, \ie
\begin{equation} \label{eq:sr_cond}
	M \, \mb \sim \mathcal{O}(1) \, ,
\end{equation}
where $M$ and $\mb$ are the mass of the black hole and boson, respectively\footnote{We use natural units throughout this work.}. For astrophysical  black holes with $M \sim \mathcal{O}(M_\odot)$, the superradiant condition (\ie \Eq{eq:sr_cond}) requires a particle mass $\mb \sim 10^{-10}$ eV. Since no boson in the Standard Model contains a mass in this range, observations of the spin distributions of black holes have been used to search for and constrain the existence of exotic ultralight particles,  \eg axions and light $Z^\prime$s~\cite{Arvanitaki:2009fg,Arvanitaki:2010sy,Arvanitaki:2014wva,Brito:2017zvb,Brito:2017wnc,Rosa:2017ury,Zhu:2020tht,Davoudiasl:2019nlo}. While it is true that no Standard Model particle has a {\emph{bare}} mass that can induce superradiance with $\sim M_\odot$ black holes, it has been pointed out that the effective mass $m_\gamma$ acquired as photons traverse the intergalactic medium (IGM) is approximately in the ideal range~\cite{Pani:2013hpa,Conlon:2017hhi}. For a non-relativistic, non-degenerate plasma comprised of electrons and ions, the plasma mass is given by
\begin{equation}\label{eq:plasma_mass}
	m_\gamma \simeq \omega_p = \sqrt{\frac{4\pi \alpha n_e}{m_e}} \, ,
\end{equation}
where $n_e$ and $m_e$ are the number density and mass of electrons, $\omega_p$ is the plasma frequency, and $\alpha$ the fine structure constant. Thus, an approximately bare black hole traversing the IGM may, under certain circumstances, undergo superradiant growth of a photon cloud. In the case of a non-interacting Proca field, superradiance is expected to efficiently extract an $\mathcal{O}(1)$ fraction of the black hole's rotational energy~\cite{Brito:2014wla,East:2017ovw}\footnote{It it worth noting that in the absence of interactions, the appearance of the superradiant instability appears (at least thus far) to be robust against back reactions on the metric~\cite{East:2013mfa}, gravitational wave emission, and accretion~\cite{Brito:2014wla}.}; in the case of the photon, non-linearities are expected to saturate the growth and complicate the evolution of the superradiant system~\cite{Fukuda:2019ewf} -- however, the extent to which these interactions inhibit the extraction of rotational energy, and influence the radiative flux and spectrum of the black hole-plasma system, remains largely unknown.

The question of whether photon superradiance can occur (or has occurred) is unclear.  In order to robustly access the viability and frequency of such events, detailed studies must be done on \eg the formation of bare rapidly spinning black holes, their environments at small radii, and the sensitivity of the superradiant event to perturbations in the radial and angular plasma frequency. A first attempt at addressing the latter point has been made in~\cite{Dima:2020rzg}, however many open questions still remain. While all of aforementioned are interesting in their own right, they require dedicated studies in and of themselves, and thus we consider them to be beyond the scope of the current work. It is worth emphasizing, however, that the stipulation of a sizable population of bare black holes is not entirely unwarranted. For instance, black hole mergers with antiparallel spins lying in the orbital plane can generate so-called `superkicks', which cause the black hole remanent to recoil against the gravitational wave emission with velocities $\sim \mathcal{O}(5000) \,$ km/s~\cite{Campanelli:2007cga,Brugmann:2007zj,Healy:2008js,Lousto:2010xk,Lousto:2011kp,Gerosa:2014gja,Gerosa:2018qay,Sperhake:2019wwo}; even larger natal kicks can be generated by hyperbolic encounters and ultra-relativistic collisions, with velocities potentially reaching $\sim 0.1 \, c$~\cite{Sperhake:2010uv}. Black holes formed with such velocities would be unable to efficiently accrete and could easily escape from their host galaxies. Similarly, the existence of a  population of highly spinning black holes which satisfy the superradiance condition $\Omega>1.5 \, {\rm Hz }\left(\frac{m_b}{10^{-13} \rm eV}\right) $ seems plausible (see also \cite{DeLuca:2020qqa,Brito:2017zvb,Brito:2017wnc,Zhu:2020tht} for further analysis of populations of highly spinning black holes in the context of dark matter models).

Hence, in what follows we start from a position in which we assume that the conditions allowing for photon superradiance are satisfied at some point in the late Universe, for a time period of at least $t \sim \mathcal{O}(1)$ minutes (which, as shown below, is approximately the timescale that would be required for a light vector boson to remove the entirety of the black hole spin), and investigate various mechanisms that could be responsible for prematurely halting the growth of the photon cloud. Specifically, we focus on processes that can directly alter the dispersion relation (\ie modifications coming from thermal correction and the presence of strong coherent electric fields), processes which directly modify the plasma mass via particle production (either via $e^\pm$ pair production, or via ionization of the neutral background), and various scattering processes.  It seems most likely that the large electric fields produced by the superradiant photons will damp the polarization tensor and allow photons to free stream, {see e.g. \cite{Cardoso:2020nst}}. Should this not be the case, Compton scattering and synchrotron cooling will be sufficiently fast so as to quench the growth long before a significant fraction of energy can be extracted. 

We emphasize that understanding the vast array of processes capable of quenching the growth of the photon cloud has value far beyond photon superradiance. Other exotic particle may acquire an effective mass via interactions with massive fermions or couple to electromagnetism in a similar manner, and thus may experience one, or many, of these quenching mechanisms. A clear example would a be a light $Z^\prime$ which couples to the Standard Model via kinetic mixing. Thus, in addition to the intended application toward photon superradiance, this work should be viewed as a guide to understanding how the superradiant growth may be quenched, or equivalently, how superradiant constraints may be evaded.

The format of this manuscript is as follows. We begin in \Sec{sec:vecSR} by reviewing the basics of superradiance in the context of a light vector boson. In \Sec{sec:plasmaM} we discuss the origin of the effective mass generated by the ambient plasma, and various processes which may back-react, changing the mass and subsequently slowing or halting the superradiant growth. We then discuss in \Sec{sec:scatter} the importance of various scattering processes that can become important after the photon cloud has extracted only a tiny fraction of the available rotational energy. We conclude in \Sec{sec:con}.

%%%%%%%%%%%%%
\section{Vector superradiance}\label{sec:vecSR}
%%%%%%%%%%%%%
Throughout this work we focus on spinning black holes embedded in a spatially uniform plasma, characterized by a plasma frequency $\omega_p$.  In the event that the mass of surrounding material is negligible relative to that of the black hole, the spacetime is uniquely characterized by the Kerr metric, defined solely by two parameters: the mass $M$, and dimensionless spin $\tilde{a} \equiv J / M^2$, with $J$ the angular momentum of the black hole. So long as the background metric is slowly varying compared to $\omega_p^{-1}$ and the density gradient is small compared to the gravitational field, Maxwell's equations yield
\begin{equation}\label{eq:proca}
	\nabla_\sigma F^{\sigma \nu} = \omega_p^2 \, A^\nu \, ,
\end{equation}
which coincides with the Proca equation. Solving \Eq{eq:proca} in a Kerr background using separation of variables has only recently been accomplished~\cite{Frolov:2018pys,Dolan:2018dqv,Cayuso:2019ieu}; here, it was shown that the superradiant instability for a near-extremal black hole the dominant mode is maximized when $M \, \omega_p \sim 0.5$. 

Of particular importance in studying the instability is an understanding of the relevant timescale over which the bosonic field grows. It has been shown that this question is readily addressed by Fourier transforming the vector field, $A_\mu(t, \vec{x}) = \int \, d\omega \, e^{-i\omega t}\tilde{A}_\mu(\omega, \vec{x})$, decomposing the frequency $\omega$ into real and imaginary parts $\omega = \omega_R + i \, \omega_I$, and working within the slow-rotation approximation~\cite{Pani:2012bp}; the timescale over which the vector field increases an e-fold then corresponds to $\tau_{SR} = \omega_I^{-1}$. The real part of the frequency and the timescale are approximately given by~\cite{Brito:2015oca,Endlich:2016jgc,Baryakhtar:2017ngi,Baumann:2019eav}
\begin{eqnarray}
	\omega_R^2 & \simeq & \omega_p^2 \left[1 - \left( \frac{M \omega_p}{\ell + n  + S + 1}\right)^2 \right]  \\ \nonumber \\ 
	M \omega_I & \simeq & 2\gamma_{S \ell} (\tilde{a} m - 2 r_+ \omega_p)(M \omega_p)^{4\ell + 5 + 2S} \, ,
\end{eqnarray}
where $r_+$ is the horizon radius, $\ell$ the harmonic index of the mode, $n$ an integer, $S$ is the polarization, and $\gamma_{S \ell}$ a coefficient that depends on $(n, \ell, m)$. The most unstable mode is that of $S = -1$, $\ell = 1$. In the case of scalar superradiance, the cloud extends to radii 
\begin{equation}\label{eq:rcloud}
	r_{\rm cloud} \sim \frac{(\ell + n + 1)^2 \, M}{(M \,\omega_p)^2} \, ,
\end{equation}
which turns out to be comparable to the Compton wavelength of the incident boson~\cite{Arvanitaki:2010sy}. Similar results are obtained for the case of a Proca field~\cite{East:2017ovw}; in what follows we will simply approximate the radial scale as being $r_{\rm cloud} \sim \lambda_\gamma \equiv 2\pi / m_\gamma$.

For the maximally growing mode and a dimensionless spin parameter $\tilde{a} \sim 0.99$, we expect the superradiant timescale of the vector field to roughly follow~\cite{Witek:2012tr,East:2017ovw,East:2017mrj,Dolan:2018dqv,Cardoso:2018tly}
\begin{equation}
	\tau_{SR} \sim \frac{M \left( M \omega_p \right)^{-7}}{\gamma_{-1 1}}\sim 10^{-4} \, \left( \frac{M}{M_\odot}\right) \, s \, ,
\end{equation}
where $\gamma_{-1 1} \sim 20$~\cite{Pani:2012bp}. Neglecting interactions and non-linearities that may inhibit the growth of the photon field, it has been shown that typically around $\mathcal{O}(10\%)$ of the energy of a near-extremal black hole may be removed before the superradiant condition is saturated~\cite{Brito:2014wla,East:2017ovw}. If one takes as an initial condition a superradiant photon number density corresponding to one photon per unit volume (where we define the superradiant volume to be $\mathcal{V} \sim 4 \pi r_{\rm cloud}^3 / 3$, then the total timescale characterizing the growth of the photon cloud around an astrophysical black hole is typically $\sim \mathcal{O}(100) \times \tau_{SR}$.  As we will discuss below, the focus of this work will be on black holes with masses $10 \, M_\odot \lesssim M \lesssim 10^4 \, M_\odot$, implying that the superradiant condition need only be satisfied for timescales $ t \sim \mathcal{O}(1) $ minutes in order to efficiently extract energy from the black hole. 

It is perhaps important to emphasize that the end state evolution of the superradiant growth is largely model-dependent. In the case of a purely non-interacting scalar or vector field, the boson cloud will grow until $\omega \sim m \Omega$, at which point the growth will saturate and the system will begin to dissipate energy via gravitational wave emission~\cite{Brito:2017zvb,Brito:2017wnc}. In more complex models, however, self-interactions may induce pressure forces that cause a premature explosion (a so-called `bosenova')~\cite{Arvanitaki:2009fg,Arvanitaki:2010sy,Yoshino:2012kn,Arvanitaki:2014wva}. Alternatively, the growth may also be saturated by the creation of new particles~\cite{Rosa:2017ury,Ikeda:2018nhb,BlasWitte_axionSRs,Mathur:2020aqv}.  In the context of the Standard Model photon, it has been pointed out that if the photon energy density exceeds $\rho_\gamma \sim 10^{36} \, {\rm eV / cm^3}$, Schwinger pair production~\cite{schwinger1951gauge} will lead to efficient creation of $e^\pm$ pairs at a rate exceeding that of superradiant growth~\cite{Fukuda:2019ewf}. Understanding the role of scattering and particle production processes is thus of the utmost importance in determining the final effect of light bosons of spinning black holes and their environments. While this study focuses explicitly on the case of the standard model photon, some of the features discussed below may also be useful for those wishing to evade superradiant constraints on exotic light bosons.

\begin{figure}
	\centering
	\includegraphics[width=0.45\textwidth, bb=0 0 580 370]{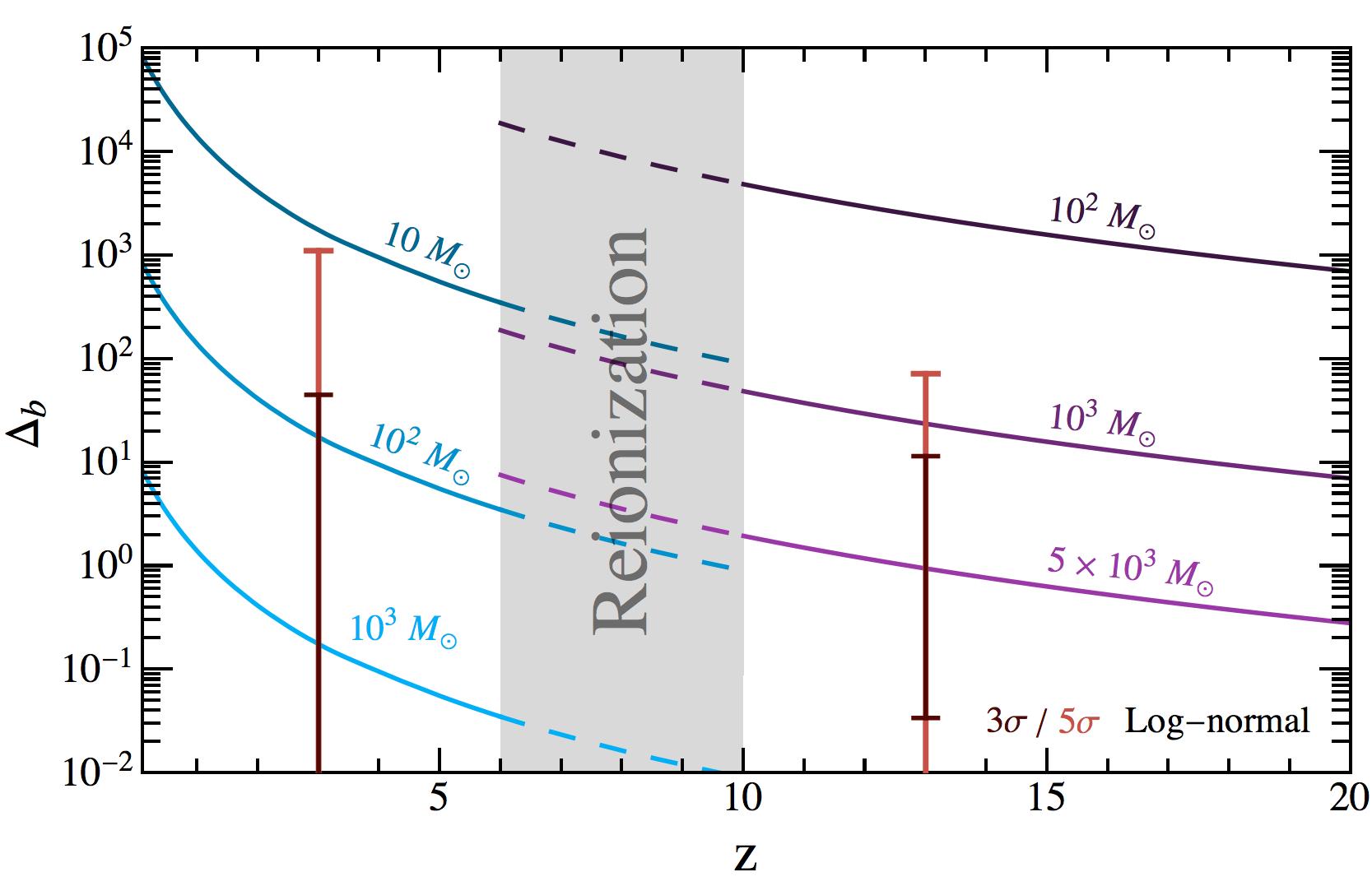}
	\caption{\label{fig:exampleSR} Baryonic overdensity $\Delta_b$  at a given redshift required to satisfy \Eq{eq:sr_cond} for various black holes masses. We overlay the $3$ and $5 \, \sigma$ two-sided confidence intervals to illustrate the potential rarity of finding such an overdensity. We highlight the approximate redshift range in which reionization is expected to take place, as the process of reionization changes the plasma mass, and consequently the black hole masses for which the  condition is satisfied, by multiple orders of magnitude. }
\end{figure}

\section{The Effective Photon Mass}\label{sec:plasmaM}

Photons propagating in a background plasma have a modified dispersion relation oweing to the presence of interactions, which for transverse (T) and longitudinal modes (L) is given by~\cite{raffelt1996stars}
\begin{eqnarray}
	\omega_T^2 & = & k^2 + \pi_T(\omega_T, k) \\
	\omega_L^2 & = & \frac{\omega_L^2}{k^2} \, \pi_L(\omega_L, k) \, ,
\end{eqnarray}
where  $\pi_{T/L}$ is the polarization tensor of each mode. In the non-degenerate and non-relativistic limit, as would be applicable for photons in the IGM, the polarization tensors can be expressed as
\begin{eqnarray}
	\pi_T & = & \omega_p^2 \\
	\pi_L & = & \omega_p^2 \frac{k^2}{\omega^2} \, ,
\end{eqnarray}
where the non-degenerate and non-relativistic formulation of the plasma mass is given by $\omega_p^2 = \sum_\alpha q^2 n_\alpha / m_\alpha$,  where $q$, $n$, and $m$ are the charge, number density, and mass of particle $\alpha$. Note that this expression is equivalent to \Eq{eq:plasma_mass} in the limit of a purely electron plasma. 

Using \Eq{eq:plasma_mass} and focusing our attention to $z \lesssim 20$, we can identify the range of black hole masses of interest for photon superradiance by analyzing the distribution of free electrons in the IGM, which we parameterize as as $n_e = \overline{n_e}(z)  \Delta = x_e \overline{n_b}(z)\,\Delta$, where $x_e$ is the free electron fraction, $\overline{n_b}$ is the mean baryon number density, and $\Delta \equiv \rho_b(\vec{x}) / \rho_b$ is the baryon overdensity. We plot in \Fig{fig:exampleSR} the baryonic over-density required to induce photon superradiance for various black hole masses as a function of redshift. We overlay $3$ and $5\sigma$ confidence intervals at $z = 3$ and $z = 13$, computed  assuming the probability distribution function of baryon over-densities can be characterized by a log-normal distribution\footnote{Notice  that the distribution of $\Delta_b$ departs from log-normal at low-$z$, in particular in the tails of the distribution~\cite{gnedin1997probing,bertschinger1998simulations}. As a consequence, the derived range may  extend a bit further than illustrated in \Fig{fig:exampleSR}.  As a word of caution, we emphasize that on sufficiently small scales one can certainly not assume the electron distribution follows that of the dark matter, as turbulent effects become increasingly important. We do not, however, expect these details to dramatically alter our conclusions. } (with a variance $\sigma^2$ computed using the non-linear dark matter power spectrum, as done in~\cite{Witte:2020rvb}), to highlight the mass range most likely to undergo superradiance. During reionization the effective photon mass changes by multiple orders of magnitude; consequently, we caution the reader that special care may be required during this epoch (the approximate redshift interval over which this process is expected to occur is highlighted in grey). \Fig{fig:exampleSR} clearly identifies the most interesting mass range for photon superradiance to be $10 \, M_\odot \lesssim M \lesssim 10^{4} \, M_\odot$.

\begin{figure}
	\centering
	\includegraphics[width=0.45\textwidth,  bb=0 0 500 630]{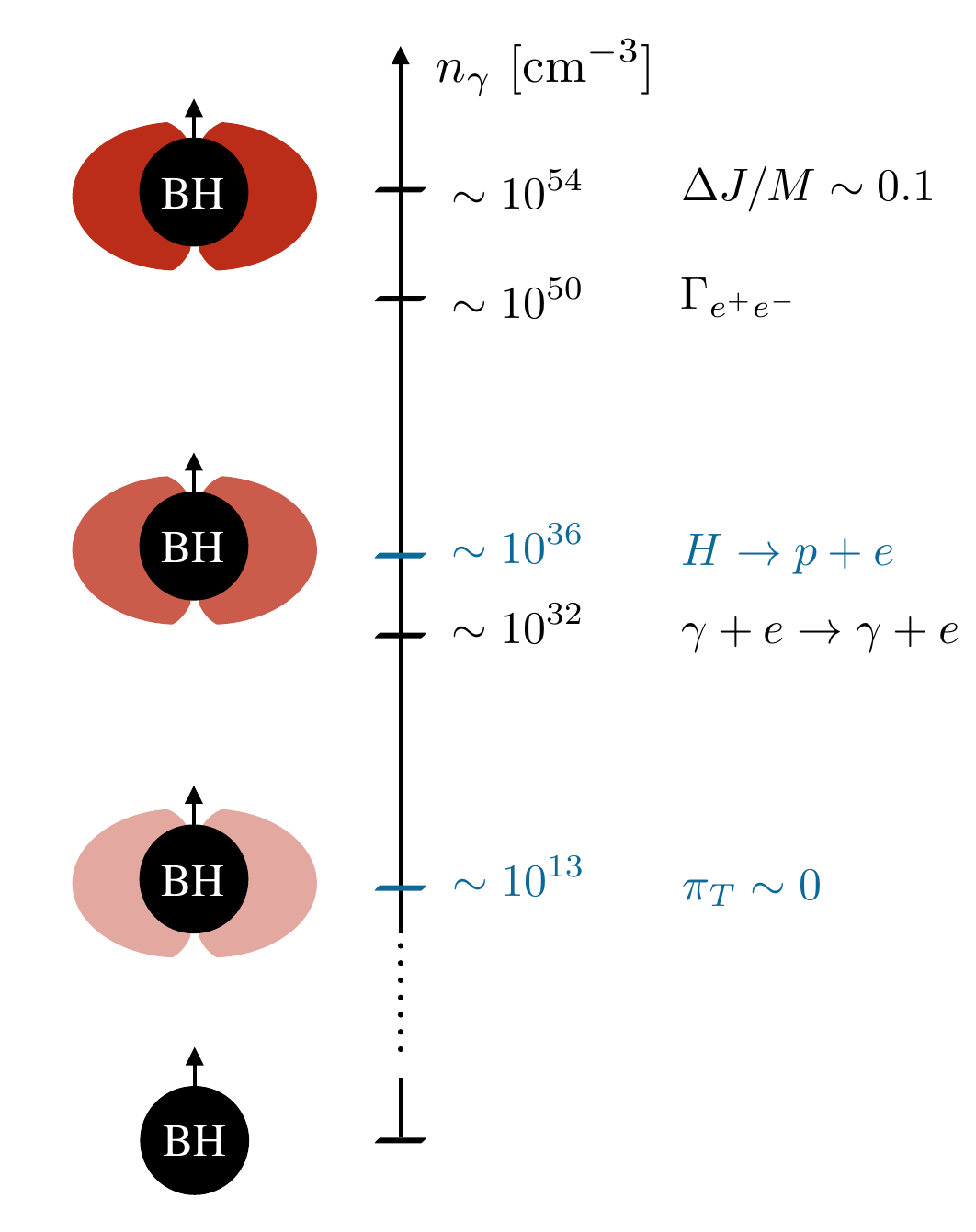}
	\caption{\label{fig:overview}  Overview of mechanisms capable of quenching photon superradiance as a function of photon number density $n_\gamma$ in units of cm$^{-3}$. Mechanisms from bottom (low number density) to top (high number density) are: the vanishing of transverse polarization tensor due to the strong electric field produced by the photon cloud, up-scattering of superradiant photons by ambient electrons, ionization via the Stark effect, Schwinger pair production,  and finally saturation of the superradiant condition. Processes shown in blue correspond to those which quench via modifications to the effective mass (other processes occur independently). All numbers are approximated using $m_\gamma \sim 10^{-13}$ eV and $M \, m_\gamma \sim 0.4$.  }
\end{figure}

In deriving the above dispersion relations, we have assumed: (1) the plasma mass is dominated by the contribution of $e^\pm$, which is true in all environments of interest, (2) the plasma is non-degenerate, (3) the plasma is non-relativistic, and (4) small external electromagnetic fields. It is straightforward to see that in the context of interest one never needs to be concerned with the degenerate plasma limit, as a modest increase in either the electron or positron number density will raise the effective photon mass and kill the superradiant condition. The leading order thermal corrections to the dispersion relations are given by ~\cite{braaten1993neutrino}
\begin{eqnarray}
	\pi_T & = & \omega_p^2 \left(1 + \frac{k^2 \, T}{\omega^2 \, m_e} \right) \\
	\pi_L & = & \omega_p^2 \left(\frac{k^2}{\omega^2} + 3 \frac{k^4 \, T}{\omega^4 \, m_e} \right) \, .
\end{eqnarray}
In order to have superradiant growth we require photons to populate quasi-bound state orbits, which only happens for non-relativistic propagating photons (\ie $k \ll \omega$). Thus we expect thermal corrections to be negligible for the photons of interest.  

Very recently, it was pointed out in~\cite{Cardoso:2020nst} that strong electric fields induced by  a large number density of coherently oscillating photons can produce significant modifications in the dispersion relation. In this case, the transverse and longitudinal polarization tensors yield~\cite{kaw1970relativistic,max1971strong}
\begin{eqnarray}
	\pi_T & = & \frac{\omega_p^2}{1 + \frac{e^2 E^2}{m_e^2 \omega^2}} \\
	\pi_L & \simeq & \frac{k^2}{\omega^2}\left( \frac{\pi \, \omega_p^2 \, m_e}{2 e E}\right)^2 \, ,
\end{eqnarray}
where in the longitudinal mode we have taken the limit of a relativistic plasma (as is the case for large electric fields). There are two interesting observations here. First, one sees that modes with $\omega < \omega_p$ are allowed to propagate when a strong electric field is present. Second, one can see that the contribution to the dispersion relation vanishes in the limit $E \gg m_e \omega$, or in the case where $E \sim \sqrt{\rho_\gamma}$ and $\omega \sim \omega_p$, $n_\gamma \gg \omega_p m_e^2$.  These terms are roughly equivalent when the photon number density is only $n_\gamma \sim 10^{14} \, {\rm cm^3}$.   When the number density greatly exceeds this threshold, the effective photon mass is expected to vanish and the photons may free stream away from the black hole. Recall that the role of the photon mass in black hole superradiance is to act as a 'mirror', to confine and reflect the radiation. Given that the photon mass at large radii will be restored, it is natural to ask whether this mechanism will truly quench the superradiance, as it seems, at least idealistically, that the reflection will always occur at some radii. While this is in some sense true, the timescale of the superradiant event is exponentially sensitive to the location of this mirror, and consequently the expectation is that the growth timescale will very quickly become larger than the accretion timescale, although it would be of value to test this hypothesis numerically. Thus it seems that the strong electric fields produced by the photon cloud itself quench the superradiant instability long before any significant amount of energy can be extracted.

We now turn our attention to alternative back-reactions on the effective photon mass that arise from the production of free electrons or positrons. This may occur if either the photon number density is sufficiently large so as to initiate pair production, or if the local medium is initially neutral and subsequently becomes ionized.

\subsection{Back-Reaction From Pair Production}

It has previously been appreciated in~\cite{Fukuda:2019ewf} that the electric fields associated with the large photon number densities obtained in the context of photon superradiance may be sufficiently high so as to trigger the Schwinger pair production \footnote{It is worth emphasizing that despite being proposed nearly a century ago~\cite{sauter1931verhalten,euler1936consequences,heisenberg2006consequences} (although, admittedly, a coherent theory of the process was not formulated until the 1950s~\cite{schwinger1951gauge}) this processes has not yet been observed; thus even an indirect observation of this phenomenon would alone be of great importance. }. This process, inducing rapid production of $e^{\pm}$ pairs, is exponentially suppressed for small electric field values. Specifically, the rate of particle production per unit volume is approximately given by
 \begin{equation} \label{eq:gammaSw}
 	\Gamma_s(n_{\gamma, sr}) \simeq \frac{m^4_e}{4\pi^3} \left(\frac{\mathcal{E}}{\mathcal{E}_c} \right)^2 \sum_{b = 1}^\infty \frac{1}{b^2} e^{-b \pi \frac{\mathcal{E}_c}{\mathcal{E}}} \, ,
 \end{equation}
where $\mathcal{E}_c = m_e^2 / \sqrt{4\pi\alpha}$ is the critical electric field and $\mathcal{E}$ the electric field induced from the superradiant photons (note that \Eq{eq:gammaSw} is dominated in the weak field limit by the $b = 1$ mode). Removing the exponential suppression of this rate requires $\mathcal{E} \sim  \sqrt{\rho_\gamma} \gtrsim 0.1\, \mathcal{E}_c$, which corresponds to a photon energy density  $\rho_{\gamma, {\rm sr}} \lesssim 10^{37} \, {\rm eV/cm^3}$. Estimating the maximal energy extracted from the black hole before the onset of pair production, assuming $m_\gamma M \sim 0.5$, we find \footnote{Note that in what follows we use $M$ as a proxy for the rotational energy of the near-extremal black holes, see e.g. \cite{DeLuca:2020bjf}.}
\begin{equation}
	\frac{E^{\rm max}_{s}}{M} \sim \frac{32 \pi^4}{3} \rho_{\gamma, c} \frac{1}{m_\gamma^3 \, M} \sim 10^{-10}\left(\frac{10^{-10} \, {\rm eV}}{m_\gamma} \right)^2 \, .
\end{equation}
Finding an effective photon mass significantly below $\sim 10^{-14}$ eV would amount to an extremely vacant void, and since it is unlikely to find black holes in such environments, we can conclude that at most $\sim 10^{-2}$ of the rotational energy can be extracted by a near extremal black hole before pair production back reacts and stops the growth.  It may be possible, however, that during the formation of the black hole, supernova winds are sufficiently strong so as to eject the ambient free electrons; this may create temporarily a void around the black hole with low electron number, allowing the the aforementioned lower bound on the effective photon mass to be evaded.

\subsection{Back-Reaction From Ionization}
An alternative back-reaction on the photon mass can occur if the black hole resides in a background of neutral hydrogen, as naturally occurs prior to reionization. Specifically, if the number density of neutral hydrogen is at least of the same order as that of the free electrons, ionization of the hydrogen could push the plasma mass above the superradiant condition. 

We consider here two possible routes in which ionization of neutral hydrogen could occur. In the first, energy is efficiently absorbed by electrons and ions via inverse bremsstrahlung absorption, and that energy is later transferred to neutral atoms via collisional processes. In the latter, we consider processes via which the bound states of hydrogen vanish due to the presence of a strong electric field, the so-called Stark effect. 

\subsubsection{Ionization from inverse bremsstrahlung absorption}

For cold low energy photons inverse bremsstrahlung (free-free) absorption is extremely efficient, and can allow for a heating of the ambient electrons and ions. Heating the electrons and ions, however, is not sufficient to guarantee ionization of the neutral hydrogen; energy must be also be efficiently transferred to the neutral atoms. Even then, one must ensure that either the photoionization or the collisional ionization timescale is short enough to significantly change the free electron fraction before the energy has been extracted from the black hole. We show explicitly below that energy transferred between the free electrons is never sufficiently fast so as to heat the neutral hydrogen. 

The dominant absorption mode for such low energy photons is inverse bremsstrahlung, with a cross section (in the limit $E_\gamma \ll T$)   given by~\cite{draine2010physics}
\begin{equation}
	\sigma_{ff} \simeq \frac{4 \pi^2 \alpha \sigma_T}{\sqrt{6\pi}} \, n_p \sqrt{\frac{m_e}{T}} \, \frac{g_{ff}(E_\gamma, T) }{T E_\gamma^2} \,  ,
\end{equation}
where $T$ is the temperature of the electrons and ions, $n_p$ is the proton number density, and $g_{ff}$ the gaunt factor, which we approximate as
 \begin{equation}\label{eq:gaunt}
 	g_{ff} \simeq 4.691 \left[1 - 0.118 \ln \left( \frac{\nu_\gamma}{10^{10} ( T / 10^4 \, {\rm K})^{3/2}} \right) \right] \, .
 \end{equation}
 Using the ambient free electron and proton densities at $z \sim 15$, and dropping the log contribution in \Eq{eq:gaunt} (note that the log term will be larger than one, and thus this assumption is overly conservative), we can write the cross section as
 \begin{equation}
 	\sigma_{ff} \sim 3.7 \times 10^{-18} \, \left( \frac{1 \, {\rm eV} }{T} \right)^{3/2} \left( \frac{10^{-13} \, {\rm eV}}{m_\gamma}\right) ^2 \, {\rm cm}^2 \, .
 \end{equation}
The relevant timescale for the free-free absorption by an electron is then given by
\begin{equation}
	\tau_{ff} = \left( \sigma_{ff} \, n_\gamma \right)^{-1} \, .
\end{equation}
In order to significantly change the energy of the electron, however, one requires many of such absorptions. We can approximate the average energy gain per electron per unit time as $\delta E  \sim m_\gamma / \tau_{ff}$. Ionizing hydrogen in conventional cosmological contexts requires the neutral hydrogen to have a temperature $T_g \gtrsim 10^{4}$ K (see \eg~\cite{McDermott:2019lch}), so we compute the photon number density necessary to produce an energy gain consistent of $1 \, {\rm eV} / \tau_{sr}$, that is the photon number density required for electrons to heat to temperatures $T\sim 10^{4} \, K$ in the superradiant timescale. Conservatively taking the temperature of the electrons and protons to be $\sim 1$ eV (this is conservative since $\sigma_{ff} \propto T^{-3/2}$) and the photon mass to be $\sim 10^{-13}$ eV,  we find the photon number density must be at least $n_\gamma \gtrsim 10^{21} \, {\rm cm}^{-3}$. This is significantly below the threshold for Schwinger pair production, and thus it seems plausible that electrons and protons will indeed become extremely hot. 

In order to ionize the medium, the energy absorbed by the electrons and protons must efficiently be shared with neutral hydrogen. The dominant process by which this happens in the IGM is via the collisional excitation or ionization of neutral hydrogen with electrons. The rate of these processes for a thermal distribution of electrons with temperature $T$ are given by~\cite{mo2010galaxy}
\begin{equation}
	\Gamma_{ci/ce} = n_e \, \left<\sigma_{ci/ce} v\right>  \, ,
\end{equation}
where the label $ci/ce$ represent collision ionization or excitation. In the case of collisional ionization, at temperatures $10^4 \, {\rm K} \, \lesssim T \lesssim 10^{8} \, {\rm K}$, this rate can be approximated as
\begin{equation}
	\Gamma_{ci} \sim \Gamma_0 \, \left(\frac{T}{T_t} \right)^{1/2} \, e^{-T_t/T} \, \left( 1 + \frac{T}{10^5 \, {\rm K}}\right)^{-1} \, n_e \, {\rm cm^3 \, s^{-1}} \, ,
\end{equation}
with $\Gamma_0 \sim 2.32 \times 10^{-8}$ and $T_t =  157809.1$ K. The collisional excitation cross section can be obtained \eg from~\cite{mendoza1983planetary}. In both cases, it is straightforward to show that the relevant timescales are at most a few orders of magnitude smaller than the Hubble time during the epochs of interest, regardless of the electron temperature. Consequently, the neutral hydrogen cannot ionize on timescales relevant to stop the growth of the photon cloud. \newline 

\subsubsection{Ionization from Stark effect}

A second possibility is that the electric field produced from the ambient superradiant photons becomes large enough that the Stark effect, \ie the splitting and shifting of atomic energy levels due to the presence of an external field, effectively removes all bound states\footnote{We thank the anonymous referee for bringing this to our attention.}. We can use the first order correction to the ground state $\epsilon_0$, given by
\begin{equation}
	\delta \epsilon_0 = - \vec{E} \cdot \vec{\mu}  \, ,
\end{equation}
where $\mu$ is the dipole moment of the atom. Using $E \sim \sqrt{m_\gamma n_\gamma}$, we find that the ground state of hydrogen vanishes when the number density of photons is approximately
\begin{equation}
	n_\gamma^{Stark} \sim 10^{35} \, \left(\frac{10^{-12} \, {\rm eV}}{m_\gamma} \right) \, {\rm cm^{-3}} \, ,
\end{equation}
amounting to only a tiny fraction of the total rotational energy of the black hole. Should photon number densities reach this level, the hydrogen will directly ionize, the photon mass will increase, and the superradiant condition of the principle mode will be violated. Thus the Stark effect offers yet another novel mechanism for quenching photon superradiance.

\section{The Role of Scattering}\label{sec:scatter}

Modifications to the plasma mass offer only one possible mechanism for halting the superradiant growth. An alternative possibility is that scattering processes may either absorb or up-scatter photons out of the quasi-bound state orbits at a rate that exceeds the energy injection from superradiance. For ultra-low energy photons, only two processes can be efficient -- free-free absorption (as discussed above), and Compton scattering. In the low energy IGM, the former process is dominant in the small $E_\gamma$ limit, however as photons are absorbed, they heat the gas and the cross section falls. Heated electrons will lose energy via Compton scattering off the photon background, and thus one can estimate the equilibrium absorption cross section by equating the rate of energy gained via free-free absorption with that emitted from Compton scattering. 

The rate of energy loss of electrons from the Compton up-scattering of a low energy photon bath with energy density $\rho_\gamma = m_\gamma n_\gamma$ is given by ~\cite{longair2011high}
\begin{equation}\label{eq:comp}
	\frac{dE_c}{dt} = \frac{4}{3}\beta^2\gamma^2 \, \sigma_T \, m_\gamma \, n_{\gamma} \, n_e \, ,
\end{equation}
while the energy gained via absorption is
\begin{equation}
	\frac{dE_{\rm abs}}{dt} = n_\gamma \, m_\gamma \, n_e \, \sigma_{ff} \, .
\end{equation}
Equating the two, one finds an equilibrium is achieved when $\sigma_{ff} \sim \gamma^2 \sigma_T$, which occurs when electrons are mildly relativistic. In order for the energy losses to balance the energy gained via superradiance, one would require an electron number density far greater than the background value, and thus free-free absorption cannot quench the photon cloud growth.

An alternative possibility for halting the growth arises when one considers the fact that the ambient electrons will be accelerated by the electromagnetic fields produced by the photon cloud. Relativistic electrons can both Compton up-scatter superradiant photons out of the cloud and emit synchrotron radiation. Assuming photons source both the electric and magnetic fields, the energy loss for synchrotron is equivalent to that of Compton scattering (given by \Eq{eq:comp}) with a suppression of $\sin^2\alpha$, $\alpha$ defining the angle between the electron velocity and magnetic field~\cite{longair2011high}. For large enough photon number densities, the motion of the electrons will be driven by electromagnetic forces, and thus Lorentz boosts can reach extremely large values $\gamma \gg 1$. 

Let us start by assuming the motion of the electrons is non-relativistic, so we can neglect the effects of the magnetic field. Here, the force generated by an oscillating electric field $E(t) = E_0 \, \cos(\omega_R \, t + \phi)$, with $E_0 \sim \sqrt{\rho_\gamma} = \sqrt{m_\gamma \, n_\gamma}$ is simply $F =dp/dt = d (\gamma m_e v)/dt = q E$.  Without loss of generality lets take the phase $\phi = \pi/2$, so that 
\begin{equation}
	\gamma(t) = \sqrt{1 + \frac{n_\gamma m_\gamma}{m_e^2 \omega_R^2} \cos^2(\omega_R t)} \, .
\end{equation}
 Assuming $n_\gamma m_\gamma / (m_e \omega_R)^2 \gg 1$ (\ie electrons are typically relativistic), and the fact that $\omega_R \sim m_\gamma$, this leads to a time averaged Lorentz boost $\left< \gamma \right> \sim \sqrt{n_\gamma m_\gamma} / (m_e m_\gamma)$.  

This energy is extracted from the electric field produced from the photons, and thus amounts to absorption of the photon field itself. The energy injected into the photon field is given by
\begin{equation}
	\frac{dE_\gamma}{dt} = m_\gamma \frac{d n_\gamma}{dt} = m_\gamma n_\gamma \frac{2}{\tau_{sr}} \, .
\end{equation}
Looking for an equilibrium solution $dE_e/dt = dE_\gamma/dt$, we find
\begin{equation}
	n_\gamma^{eq} = 6\pi\alpha \frac{m_e}{m_\gamma}\frac{1}{\sigma_T \, \tau_{sr}} \, .
\end{equation}
Using $m_\gamma \, M \equiv \tilde{\mu}$, we find $n_\gamma^{eq} \sim 3.5 \times 10^{32} \, {\rm cm}^{-3}$. This implies a typical Lorentz boost of $\left< \gamma \right> \sim 10^{10}$ for $m_\gamma \sim 10^{-13}$ eV, implying the up-scattered photons escape with energies $\sim \gamma^2 m_\gamma \sim 10$ MeV,  and stop the growth. The luminosity of these events is roughly
\begin{equation}
	L = \frac{dE_e}{dt} \mathcal{V} \sim \frac{32 \, \pi^3 \, n_\gamma^2 \, \sigma_T}{9 \, \alpha \, m_e \, m_\gamma} \, .
\end{equation}
While we had originally neglected the role of the magnetic field, once the electrons become relativistic the magnetic force will become important, modifying the trajectory of the electrons and inducing synchrotron radiation. The energy loss rage from relativistic electrons emitting synchrotron radiation is given by
\begin{equation}
	\frac{dE_{\rm syn}}{dt} = \frac{4}{3} \, \beta^2 \, \gamma^2 \sigma_T \, u_{B} \, n_e \,  \sin^2\alpha \, ,
\end{equation}
where $u_B$ is the energy density in the magnetic field and $\alpha$ the angle between the magnetic field and the motion of the electrons~\cite{longair2011high}. For $u_B \sim u_E \sim n_\gamma \, m_\gamma$, as is the case for electromagnetic fields generated by a coherent field of photons, the energy radiated in synchrotron is approximately equivalent that lost via inverse Compton scattering. For each process, we estimate the luminosity to be
\begin{equation}
	L \sim 5 \times 10^{48}\left( \frac{10^{-13} \, {\rm eV}}{m_\gamma} \right) \frac{{\rm eV}}{\rm s} \,  .
\end{equation}
The Compton up-scattered photons will be sitting in the middle of the x-ray band, and will have mean free paths that are extremely large, allowing them to effectively free stream today. 

In the case of synchrotron emission, most of the energy is located near $E_\gamma \sim \gamma^2 \omega_g$, where $\omega_g = e B / m_e$ is the gyrofrequency. This corresponds to super energetic gamma-rays with energies $\sim 10^8$ GeV. Such high energy photons will pair produce $e^\pm$ and create electromagnetic showers spanning a wide array of frequencies. Due to the complicated nature of this process, we do not attempt to compute the final state spectrum.

\section{Conclusions} \label{sec:con}

In this work, we have investigated the extent to which the Standard Model photon can induce superradiance in astrophysical black holes, with the specific intent of identifying the mechanism responsible for quenching the growth of the superradiant cloud, and any subsequent astrophysical signatures that may allow for the observation of this phenomenon.  A summary of the relevant effects capable of quenching superradiance is shown in \Fig{fig:overview}. For each quenching mechanism, we highlight the approximate number density at which the superradiant growth is expected to stop (computed assuming $m_\gamma \sim 10^{-13}\,{\rm eV}$ and $M \, m_\gamma \sim 0.4$), and whether the process relies on modifying the effective photon mass (such processes are highlighted in blue). 

In this manuscript we outlined the various processes which could directly alter the dispersion relation or plasma frequency itself, as minor modifications could either invalidate the superradiant condition or lead to an exponential increase in the superradiant timescale; these processes include thermal corrections to the plasma mass, modifications to the dispersion relation from strong electric fields, Schwinger pair production, and a local ionization of the ambient neutral hydrogen. Particle production via the Schwinger mechanism becomes efficient only after the photon field has grown to energy densities $\rho_\gamma \sim 10^{37} \, {\rm eV / cm^3}$, at which point a non-negligible fraction of the energy would have been extracted from the black hole (potentially as much as $\sim 10^{-2} \, M$). We investigate two possible mechanisms for ionizing the local hydrogen. in the first, ionization from the absorption of the low energy photon, we show that ionization can never occur on sufficiently short timescales so as  to modify the growth of the photon cloud, owing to the fact that the e-H scattering timescale is far greater than that of superradiance. The second mechanism, however, arises from Stark effect, in which the energy levels of neutral hydrogen can effectively be removed in the presence of a strong electric field -- this process occurs for photon number densities $n_\gamma \sim 10^{35} \, {\rm cm^{-3}}$. The most important process identified arises from a modification to the dispersion relation from the electric field produced by the superradiant photons themselves (recently studied in~\cite{Cardoso:2020nst}); this effect occurs for photon number densities as small as $n_\gamma \sim 10^{14} \, {\rm cm^{-3}}$, long before a significant fraction of energy has been extracted.

We have also considered the role that particle scattering processes play in inhibiting the growth of the photon cloud. We have shown that  the large electromagnetic fields induced by the growing  cloud induce rapid oscillations in the ambient free electrons; since the synchrotron and Compton cooling rate of relativistic electrons is proportional to $\gamma^2$, the system is capable of reaching an equilibrium in which the rotational energy extracted from the black hole directly balances the energy radiated via cooling processes. This equilibrium occurs for electron number densities $n_\gamma \sim 10^{32} \, {\rm cm^{-3}}$, long before the onset of pair production, but after modifications to the dispersion relation from the existence of the electric field have become important. The Compton up-scattering of the superradiant photons would produce a flux of $\sim$MeV scale x-rays, while synchrotron cooling generates ultra-energetic gamma rays (potentially exceeding $\sim 10^5$ TeV), resulting in large electromagnetic showers in the nearby medium.

Black hole superradiance has been quite successful in excluding the existence of ultralight non-interacting bosons. These bounds, however, can be incredibly sensitive to the details of the underlying model, and various mechanisms have been proposed in order to evade such constraints (\eg see~\cite{Fukuda:2019ewf,Mathur:2020aqv,BlasWitte_axionSR,Cardoso:2020nst}). Since the photon represents the quintessential example of an ultralight interacting boson, understanding {\emph{all}} physical processes that are capable of inhibiting superradiant growth of the photon may yield insight into novel mechanisms that can be used to evade superradiant constraints.  A clear example where this may be relevant is that of the dark photon; here, the existence of a bare mass will allow number densities to exceed the threshold identified for the Standard Model photon, but electromagnetic interactions with the ambient plasma may still quench the growth -- we leave a detailed study of this model to future work.

Our analysis has identified a vast array of processes which strongly inhibit the growth of superradiant photon clouds. These may arise either from modifications to the effective mass (via direct modifications to the dispersion relation or the plasma mass itself) or from particle scattering processes.  Despite the fact that photon superradiance seems incapable of efficiently extracting energy from black holes, understanding the quenching mechanisms for this mechanism may yield valued insight into the superradiant growth using realistic models of weakly-coupled ultralight bosons.

%These exciting possibilities are to be taken with a grain of salt, and call for a more thorough study of the approximations we mentioned throughout the text. Similarly, one could try to use population studies of spinning black holes, as those considered in e.g.  \cite{DeLuca:2020qqa,Brito:2017zvb,Brito:2017wnc,Zhu:2020tht}, and try to elucidate the scenarios where the BHs of interest may be found in the IGM, e.g. by `superkicks'~\cite{Campanelli:2007cga,Brugmann:2007zj,Healy:2008js,Lousto:2010xk,Lousto:2011kp,Gerosa:2014gja,Gerosa:2018qay,Sperhake:2019wwo}. 
%We hope to explore these possibilities in the future. 

~\\ \noindent{\it \bf Acknowledgments:} 
The authors thank Vitor Cardoso, Juli\'an Mu\~noz and Paolo Pani for their comments. SJW acknowledges support under the Juan de la Cierva Formaci\'{o}n Fellowship.

\appendix
\bibliography{biblio}

\begin{thebibliography}{64}%
\makeatletter
\providecommand \@ifxundefined [1]{%
 \@ifx{#1\undefined}
}%
\providecommand \@ifnum [1]{%
 \ifnum #1\expandafter \@firstoftwo
 \else \expandafter \@secondoftwo
 \fi
}%
\providecommand \@ifx [1]{%
 \ifx #1\expandafter \@firstoftwo
 \else \expandafter \@secondoftwo
 \fi
}%
\providecommand \natexlab [1]{#1}%
\providecommand \enquote  [1]{``#1''}%
\providecommand \bibnamefont  [1]{#1}%
\providecommand \bibfnamefont [1]{#1}%
\providecommand \citenamefont [1]{#1}%
\providecommand \href@noop [0]{\@secondoftwo}%
\providecommand \href [0]{\begingroup \@sanitize@url \@href}%
\providecommand \@href[1]{\@@startlink{#1}\@@href}%
\providecommand \@@href[1]{\endgroup#1\@@endlink}%
\providecommand \@sanitize@url [0]{\catcode `\\12\catcode `\$12\catcode
  `\&12\catcode `\#12\catcode `\^12\catcode `\_12\catcode `\%12\relax}%
\providecommand \@@startlink[1]{}%
\providecommand \@@endlink[0]{}%
\providecommand \url  [0]{\begingroup\@sanitize@url \@url }%
\providecommand \@url [1]{\endgroup\@href {#1}{\urlprefix }}%
\providecommand \urlprefix  [0]{URL }%
\providecommand \Eprint [0]{\href }%
\providecommand \doibase [0]{http://dx.doi.org/}%
\providecommand \selectlanguage [0]{\@gobble}%
\providecommand \bibinfo  [0]{\@secondoftwo}%
\providecommand \bibfield  [0]{\@secondoftwo}%
\providecommand \translation [1]{[#1]}%
\providecommand \BibitemOpen [0]{}%
\providecommand \bibitemStop [0]{}%
\providecommand \bibitemNoStop [0]{.\EOS\space}%
\providecommand \EOS [0]{\spacefactor3000\relax}%
\providecommand \BibitemShut  [1]{\csname bibitem#1\endcsname}%
\let\auto@bib@innerbib\@empty
%</preamble>
\bibitem [{\citenamefont {Zel'Dovich}(1971)}]{zel1971generation}%
  \BibitemOpen
  \bibfield  {author} {\bibinfo {author} {\bibfnamefont {Y.~B.}\ \bibnamefont
  {Zel'Dovich}},\ }\href@noop {} {\bibfield  {journal} {\bibinfo  {journal}
  {JETPL}\ }\textbf {\bibinfo {volume} {14}},\ \bibinfo {pages} {180} (\bibinfo
  {year} {1971})}\BibitemShut {NoStop}%
\bibitem [{\citenamefont {Press}\ and\ \citenamefont
  {Teukolsky}(1972)}]{Press:1972zz}%
  \BibitemOpen
  \bibfield  {author} {\bibinfo {author} {\bibfnamefont {W.~H.}\ \bibnamefont
  {Press}}\ and\ \bibinfo {author} {\bibfnamefont {S.~A.}\ \bibnamefont
  {Teukolsky}},\ }\href {\doibase 10.1038/238211a0} {\bibfield  {journal}
  {\bibinfo  {journal} {Nature}\ }\textbf {\bibinfo {volume} {238}},\ \bibinfo
  {pages} {211} (\bibinfo {year} {1972})}\BibitemShut {NoStop}%
\bibitem [{\citenamefont {Teukolsky}\ and\ \citenamefont
  {Press}(1974)}]{Teukolsky:1974yv}%
  \BibitemOpen
  \bibfield  {author} {\bibinfo {author} {\bibfnamefont {S.}~\bibnamefont
  {Teukolsky}}\ and\ \bibinfo {author} {\bibfnamefont {W.}~\bibnamefont
  {Press}},\ }\href {\doibase 10.1086/153180} {\bibfield  {journal} {\bibinfo
  {journal} {Astrophys. J.}\ }\textbf {\bibinfo {volume} {193}},\ \bibinfo
  {pages} {443} (\bibinfo {year} {1974})}\BibitemShut {NoStop}%
\bibitem [{\citenamefont {Cardoso}\ \emph {et~al.}(2004)\citenamefont
  {Cardoso}, \citenamefont {Dias}, \citenamefont {Lemos},\ and\ \citenamefont
  {Yoshida}}]{Cardoso:2004nk}%
  \BibitemOpen
  \bibfield  {author} {\bibinfo {author} {\bibfnamefont {V.}~\bibnamefont
  {Cardoso}}, \bibinfo {author} {\bibfnamefont {O.~J.}\ \bibnamefont {Dias}},
  \bibinfo {author} {\bibfnamefont {J.~P.}\ \bibnamefont {Lemos}}, \ and\
  \bibinfo {author} {\bibfnamefont {S.}~\bibnamefont {Yoshida}},\ }\href
  {\doibase 10.1103/PhysRevD.70.049903} {\bibfield  {journal} {\bibinfo
  {journal} {Phys. Rev. D}\ }\textbf {\bibinfo {volume} {70}},\ \bibinfo
  {pages} {044039} (\bibinfo {year} {2004})},\ \bibinfo {note} {[Erratum:
  Phys.Rev.D 70, 049903 (2004)]},\ \Eprint
  {http://arxiv.org/abs/hep-th/0404096} {arXiv:hep-th/0404096} \BibitemShut
  {NoStop}%
\bibitem [{\citenamefont {Damour}\ \emph {et~al.}(1976)\citenamefont {Damour},
  \citenamefont {Deruelle},\ and\ \citenamefont {Ruffini}}]{Damour:1976kh}%
  \BibitemOpen
  \bibfield  {author} {\bibinfo {author} {\bibfnamefont {T.}~\bibnamefont
  {Damour}}, \bibinfo {author} {\bibfnamefont {N.}~\bibnamefont {Deruelle}}, \
  and\ \bibinfo {author} {\bibfnamefont {R.}~\bibnamefont {Ruffini}},\ }\href
  {\doibase 10.1007/BF02725534} {\bibfield  {journal} {\bibinfo  {journal}
  {Lett. Nuovo Cim.}\ }\textbf {\bibinfo {volume} {15}},\ \bibinfo {pages}
  {257} (\bibinfo {year} {1976})}\BibitemShut {NoStop}%
\bibitem [{\citenamefont {Detweiler}(1980)}]{detweiler1980klein}%
  \BibitemOpen
  \bibfield  {author} {\bibinfo {author} {\bibfnamefont {S.}~\bibnamefont
  {Detweiler}},\ }\href@noop {} {\bibfield  {journal} {\bibinfo  {journal}
  {Physical Review D}\ }\textbf {\bibinfo {volume} {22}},\ \bibinfo {pages}
  {2323} (\bibinfo {year} {1980})}\BibitemShut {NoStop}%
\bibitem [{\citenamefont {Brito}\ \emph
  {et~al.}(2015{\natexlab{a}})\citenamefont {Brito}, \citenamefont {Cardoso},\
  and\ \citenamefont {Pani}}]{Brito:2015oca}%
  \BibitemOpen
  \bibfield  {author} {\bibinfo {author} {\bibfnamefont {R.}~\bibnamefont
  {Brito}}, \bibinfo {author} {\bibfnamefont {V.}~\bibnamefont {Cardoso}}, \
  and\ \bibinfo {author} {\bibfnamefont {P.}~\bibnamefont {Pani}},\ }\href@noop
  {} {\emph {\bibinfo {title} {Superradiance}}},\ Vol.\ \bibinfo {volume}
  {906}\ (\bibinfo  {publisher} {Springer},\ \bibinfo {year} {2015})\ pp.\
  \bibinfo {pages} {1501--06570},\ \Eprint {http://arxiv.org/abs/1501.06570}
  {arXiv:1501.06570 [gr-qc]} \BibitemShut {NoStop}%
\bibitem [{\citenamefont {Witek}\ \emph {et~al.}(2013)\citenamefont {Witek},
  \citenamefont {Cardoso}, \citenamefont {Ishibashi},\ and\ \citenamefont
  {Sperhake}}]{Witek:2012tr}%
  \BibitemOpen
  \bibfield  {author} {\bibinfo {author} {\bibfnamefont {H.}~\bibnamefont
  {Witek}}, \bibinfo {author} {\bibfnamefont {V.}~\bibnamefont {Cardoso}},
  \bibinfo {author} {\bibfnamefont {A.}~\bibnamefont {Ishibashi}}, \ and\
  \bibinfo {author} {\bibfnamefont {U.}~\bibnamefont {Sperhake}},\ }\href
  {\doibase 10.1103/PhysRevD.87.043513} {\bibfield  {journal} {\bibinfo
  {journal} {Phys. Rev. D}\ }\textbf {\bibinfo {volume} {87}},\ \bibinfo
  {pages} {043513} (\bibinfo {year} {2013})},\ \Eprint
  {http://arxiv.org/abs/1212.0551} {arXiv:1212.0551 [gr-qc]} \BibitemShut
  {NoStop}%
\bibitem [{\citenamefont {Arvanitaki}\ \emph {et~al.}(2010)\citenamefont
  {Arvanitaki}, \citenamefont {Dimopoulos}, \citenamefont {Dubovsky},
  \citenamefont {Kaloper},\ and\ \citenamefont
  {March-Russell}}]{Arvanitaki:2009fg}%
  \BibitemOpen
  \bibfield  {author} {\bibinfo {author} {\bibfnamefont {A.}~\bibnamefont
  {Arvanitaki}}, \bibinfo {author} {\bibfnamefont {S.}~\bibnamefont
  {Dimopoulos}}, \bibinfo {author} {\bibfnamefont {S.}~\bibnamefont
  {Dubovsky}}, \bibinfo {author} {\bibfnamefont {N.}~\bibnamefont {Kaloper}}, \
  and\ \bibinfo {author} {\bibfnamefont {J.}~\bibnamefont {March-Russell}},\
  }\href {\doibase 10.1103/PhysRevD.81.123530} {\bibfield  {journal} {\bibinfo
  {journal} {Phys. Rev. D}\ }\textbf {\bibinfo {volume} {81}},\ \bibinfo
  {pages} {123530} (\bibinfo {year} {2010})},\ \Eprint
  {http://arxiv.org/abs/0905.4720} {arXiv:0905.4720 [hep-th]} \BibitemShut
  {NoStop}%
\bibitem [{\citenamefont {Arvanitaki}\ and\ \citenamefont
  {Dubovsky}(2011)}]{Arvanitaki:2010sy}%
  \BibitemOpen
  \bibfield  {author} {\bibinfo {author} {\bibfnamefont {A.}~\bibnamefont
  {Arvanitaki}}\ and\ \bibinfo {author} {\bibfnamefont {S.}~\bibnamefont
  {Dubovsky}},\ }\href {\doibase 10.1103/PhysRevD.83.044026} {\bibfield
  {journal} {\bibinfo  {journal} {Phys. Rev. D}\ }\textbf {\bibinfo {volume}
  {83}},\ \bibinfo {pages} {044026} (\bibinfo {year} {2011})},\ \Eprint
  {http://arxiv.org/abs/1004.3558} {arXiv:1004.3558 [hep-th]} \BibitemShut
  {NoStop}%
\bibitem [{\citenamefont {Arvanitaki}\ \emph {et~al.}(2015)\citenamefont
  {Arvanitaki}, \citenamefont {Baryakhtar},\ and\ \citenamefont
  {Huang}}]{Arvanitaki:2014wva}%
  \BibitemOpen
  \bibfield  {author} {\bibinfo {author} {\bibfnamefont {A.}~\bibnamefont
  {Arvanitaki}}, \bibinfo {author} {\bibfnamefont {M.}~\bibnamefont
  {Baryakhtar}}, \ and\ \bibinfo {author} {\bibfnamefont {X.}~\bibnamefont
  {Huang}},\ }\href {\doibase 10.1103/PhysRevD.91.084011} {\bibfield  {journal}
  {\bibinfo  {journal} {Phys. Rev. D}\ }\textbf {\bibinfo {volume} {91}},\
  \bibinfo {pages} {084011} (\bibinfo {year} {2015})},\ \Eprint
  {http://arxiv.org/abs/1411.2263} {arXiv:1411.2263 [hep-ph]} \BibitemShut
  {NoStop}%
\bibitem [{\citenamefont {Brito}\ \emph
  {et~al.}(2017{\natexlab{a}})\citenamefont {Brito}, \citenamefont {Ghosh},
  \citenamefont {Barausse}, \citenamefont {Berti}, \citenamefont {Cardoso},
  \citenamefont {Dvorkin}, \citenamefont {Klein},\ and\ \citenamefont
  {Pani}}]{Brito:2017zvb}%
  \BibitemOpen
  \bibfield  {author} {\bibinfo {author} {\bibfnamefont {R.}~\bibnamefont
  {Brito}}, \bibinfo {author} {\bibfnamefont {S.}~\bibnamefont {Ghosh}},
  \bibinfo {author} {\bibfnamefont {E.}~\bibnamefont {Barausse}}, \bibinfo
  {author} {\bibfnamefont {E.}~\bibnamefont {Berti}}, \bibinfo {author}
  {\bibfnamefont {V.}~\bibnamefont {Cardoso}}, \bibinfo {author} {\bibfnamefont
  {I.}~\bibnamefont {Dvorkin}}, \bibinfo {author} {\bibfnamefont
  {A.}~\bibnamefont {Klein}}, \ and\ \bibinfo {author} {\bibfnamefont
  {P.}~\bibnamefont {Pani}},\ }\href {\doibase 10.1103/PhysRevD.96.064050}
  {\bibfield  {journal} {\bibinfo  {journal} {Phys. Rev. D}\ }\textbf {\bibinfo
  {volume} {96}},\ \bibinfo {pages} {064050} (\bibinfo {year}
  {2017}{\natexlab{a}})},\ \Eprint {http://arxiv.org/abs/1706.06311}
  {arXiv:1706.06311 [gr-qc]} \BibitemShut {NoStop}%
\bibitem [{\citenamefont {Brito}\ \emph
  {et~al.}(2017{\natexlab{b}})\citenamefont {Brito}, \citenamefont {Ghosh},
  \citenamefont {Barausse}, \citenamefont {Berti}, \citenamefont {Cardoso},
  \citenamefont {Dvorkin}, \citenamefont {Klein},\ and\ \citenamefont
  {Pani}}]{Brito:2017wnc}%
  \BibitemOpen
  \bibfield  {author} {\bibinfo {author} {\bibfnamefont {R.}~\bibnamefont
  {Brito}}, \bibinfo {author} {\bibfnamefont {S.}~\bibnamefont {Ghosh}},
  \bibinfo {author} {\bibfnamefont {E.}~\bibnamefont {Barausse}}, \bibinfo
  {author} {\bibfnamefont {E.}~\bibnamefont {Berti}}, \bibinfo {author}
  {\bibfnamefont {V.}~\bibnamefont {Cardoso}}, \bibinfo {author} {\bibfnamefont
  {I.}~\bibnamefont {Dvorkin}}, \bibinfo {author} {\bibfnamefont
  {A.}~\bibnamefont {Klein}}, \ and\ \bibinfo {author} {\bibfnamefont
  {P.}~\bibnamefont {Pani}},\ }\href {\doibase 10.1103/PhysRevLett.119.131101}
  {\bibfield  {journal} {\bibinfo  {journal} {Phys. Rev. Lett.}\ }\textbf
  {\bibinfo {volume} {119}},\ \bibinfo {pages} {131101} (\bibinfo {year}
  {2017}{\natexlab{b}})},\ \Eprint {http://arxiv.org/abs/1706.05097}
  {arXiv:1706.05097 [gr-qc]} \BibitemShut {NoStop}%
\bibitem [{\citenamefont {Rosa}\ and\ \citenamefont
  {Kephart}(2018)}]{Rosa:2017ury}%
  \BibitemOpen
  \bibfield  {author} {\bibinfo {author} {\bibfnamefont {J.~G.}\ \bibnamefont
  {Rosa}}\ and\ \bibinfo {author} {\bibfnamefont {T.~W.}\ \bibnamefont
  {Kephart}},\ }\href {\doibase 10.1103/PhysRevLett.120.231102} {\bibfield
  {journal} {\bibinfo  {journal} {Phys. Rev. Lett.}\ }\textbf {\bibinfo
  {volume} {120}},\ \bibinfo {pages} {231102} (\bibinfo {year} {2018})},\
  \Eprint {http://arxiv.org/abs/1709.06581} {arXiv:1709.06581 [gr-qc]}
  \BibitemShut {NoStop}%
\bibitem [{\citenamefont {Zhu}\ \emph {et~al.}(2020)\citenamefont {Zhu},
  \citenamefont {Baryakhtar}, \citenamefont {Papa}, \citenamefont {Tsuna},
  \citenamefont {Kawanaka},\ and\ \citenamefont {Eggenstein}}]{Zhu:2020tht}%
  \BibitemOpen
  \bibfield  {author} {\bibinfo {author} {\bibfnamefont {S.~J.}\ \bibnamefont
  {Zhu}}, \bibinfo {author} {\bibfnamefont {M.}~\bibnamefont {Baryakhtar}},
  \bibinfo {author} {\bibfnamefont {M.~A.}\ \bibnamefont {Papa}}, \bibinfo
  {author} {\bibfnamefont {D.}~\bibnamefont {Tsuna}}, \bibinfo {author}
  {\bibfnamefont {N.}~\bibnamefont {Kawanaka}}, \ and\ \bibinfo {author}
  {\bibfnamefont {H.-B.}\ \bibnamefont {Eggenstein}},\ }\href@noop {} {\
  (\bibinfo {year} {2020})},\ \Eprint {http://arxiv.org/abs/2003.03359}
  {arXiv:2003.03359 [gr-qc]} \BibitemShut {NoStop}%
\bibitem [{\citenamefont {Davoudiasl}\ and\ \citenamefont
  {Denton}(2019)}]{Davoudiasl:2019nlo}%
  \BibitemOpen
  \bibfield  {author} {\bibinfo {author} {\bibfnamefont {H.}~\bibnamefont
  {Davoudiasl}}\ and\ \bibinfo {author} {\bibfnamefont {P.~B.}\ \bibnamefont
  {Denton}},\ }\href {\doibase 10.1103/PhysRevLett.123.021102} {\bibfield
  {journal} {\bibinfo  {journal} {Phys. Rev. Lett.}\ }\textbf {\bibinfo
  {volume} {123}},\ \bibinfo {pages} {021102} (\bibinfo {year} {2019})},\
  \Eprint {http://arxiv.org/abs/1904.09242} {arXiv:1904.09242 [astro-ph.CO]}
  \BibitemShut {NoStop}%
\bibitem [{\citenamefont {Pani}\ and\ \citenamefont
  {Loeb}(2013)}]{Pani:2013hpa}%
  \BibitemOpen
  \bibfield  {author} {\bibinfo {author} {\bibfnamefont {P.}~\bibnamefont
  {Pani}}\ and\ \bibinfo {author} {\bibfnamefont {A.}~\bibnamefont {Loeb}},\
  }\href {\doibase 10.1103/PhysRevD.88.041301} {\bibfield  {journal} {\bibinfo
  {journal} {Phys. Rev. D}\ }\textbf {\bibinfo {volume} {88}},\ \bibinfo
  {pages} {041301} (\bibinfo {year} {2013})},\ \Eprint
  {http://arxiv.org/abs/1307.5176} {arXiv:1307.5176 [astro-ph.CO]} \BibitemShut
  {NoStop}%
\bibitem [{\citenamefont {Conlon}\ and\ \citenamefont
  {Herdeiro}(2018)}]{Conlon:2017hhi}%
  \BibitemOpen
  \bibfield  {author} {\bibinfo {author} {\bibfnamefont {J.~P.}\ \bibnamefont
  {Conlon}}\ and\ \bibinfo {author} {\bibfnamefont {C.~A.}\ \bibnamefont
  {Herdeiro}},\ }\href {\doibase 10.1016/j.physletb.2018.02.073} {\bibfield
  {journal} {\bibinfo  {journal} {Phys. Lett. B}\ }\textbf {\bibinfo {volume}
  {780}},\ \bibinfo {pages} {169} (\bibinfo {year} {2018})},\ \Eprint
  {http://arxiv.org/abs/1701.02034} {arXiv:1701.02034 [astro-ph.HE]}
  \BibitemShut {NoStop}%
\bibitem [{\citenamefont {Brito}\ \emph
  {et~al.}(2015{\natexlab{b}})\citenamefont {Brito}, \citenamefont {Cardoso},\
  and\ \citenamefont {Pani}}]{Brito:2014wla}%
  \BibitemOpen
  \bibfield  {author} {\bibinfo {author} {\bibfnamefont {R.}~\bibnamefont
  {Brito}}, \bibinfo {author} {\bibfnamefont {V.}~\bibnamefont {Cardoso}}, \
  and\ \bibinfo {author} {\bibfnamefont {P.}~\bibnamefont {Pani}},\ }\href
  {\doibase 10.1088/0264-9381/32/13/134001} {\bibfield  {journal} {\bibinfo
  {journal} {Class. Quant. Grav.}\ }\textbf {\bibinfo {volume} {32}},\ \bibinfo
  {pages} {134001} (\bibinfo {year} {2015}{\natexlab{b}})},\ \Eprint
  {http://arxiv.org/abs/1411.0686} {arXiv:1411.0686 [gr-qc]} \BibitemShut
  {NoStop}%
\bibitem [{\citenamefont {East}\ and\ \citenamefont
  {Pretorius}(2017)}]{East:2017ovw}%
  \BibitemOpen
  \bibfield  {author} {\bibinfo {author} {\bibfnamefont {W.~E.}\ \bibnamefont
  {East}}\ and\ \bibinfo {author} {\bibfnamefont {F.}~\bibnamefont
  {Pretorius}},\ }\href {\doibase 10.1103/PhysRevLett.119.041101} {\bibfield
  {journal} {\bibinfo  {journal} {Phys. Rev. Lett.}\ }\textbf {\bibinfo
  {volume} {119}},\ \bibinfo {pages} {041101} (\bibinfo {year} {2017})},\
  \Eprint {http://arxiv.org/abs/1704.04791} {arXiv:1704.04791 [gr-qc]}
  \BibitemShut {NoStop}%
\bibitem [{\citenamefont {East}\ \emph {et~al.}(2014)\citenamefont {East},
  \citenamefont {Ramazanou~glu},\ and\ \citenamefont
  {Pretorius}}]{East:2013mfa}%
  \BibitemOpen
  \bibfield  {author} {\bibinfo {author} {\bibfnamefont {W.~E.}\ \bibnamefont
  {East}}, \bibinfo {author} {\bibfnamefont {F.~M.}\ \bibnamefont
  {Ramazanou~glu}}, \ and\ \bibinfo {author} {\bibfnamefont {F.}~\bibnamefont
  {Pretorius}},\ }\href {\doibase 10.1103/PhysRevD.89.061503} {\bibfield
  {journal} {\bibinfo  {journal} {Phys. Rev. D}\ }\textbf {\bibinfo {volume}
  {89}},\ \bibinfo {pages} {061503} (\bibinfo {year} {2014})},\ \Eprint
  {http://arxiv.org/abs/1312.4529} {arXiv:1312.4529 [gr-qc]} \BibitemShut
  {NoStop}%
\bibitem [{\citenamefont {Fukuda}\ and\ \citenamefont
  {Nakayama}(2020)}]{Fukuda:2019ewf}%
  \BibitemOpen
  \bibfield  {author} {\bibinfo {author} {\bibfnamefont {H.}~\bibnamefont
  {Fukuda}}\ and\ \bibinfo {author} {\bibfnamefont {K.}~\bibnamefont
  {Nakayama}},\ }\href {\doibase 10.1007/JHEP01(2020)128} {\bibfield  {journal}
  {\bibinfo  {journal} {JHEP}\ }\textbf {\bibinfo {volume} {01}},\ \bibinfo
  {pages} {128} (\bibinfo {year} {2020})},\ \Eprint
  {http://arxiv.org/abs/1910.06308} {arXiv:1910.06308 [hep-ph]} \BibitemShut
  {NoStop}%
\bibitem [{\citenamefont {Dima}\ and\ \citenamefont
  {Barausse}(2020)}]{Dima:2020rzg}%
  \BibitemOpen
  \bibfield  {author} {\bibinfo {author} {\bibfnamefont {A.}~\bibnamefont
  {Dima}}\ and\ \bibinfo {author} {\bibfnamefont {E.}~\bibnamefont
  {Barausse}},\ }\href@noop {} {\  (\bibinfo {year} {2020})},\ \Eprint
  {http://arxiv.org/abs/2001.11484} {arXiv:2001.11484 [gr-qc]} \BibitemShut
  {NoStop}%
\bibitem [{\citenamefont {Campanelli}\ \emph {et~al.}(2007)\citenamefont
  {Campanelli}, \citenamefont {Lousto}, \citenamefont {Zlochower},\ and\
  \citenamefont {Merritt}}]{Campanelli:2007cga}%
  \BibitemOpen
  \bibfield  {author} {\bibinfo {author} {\bibfnamefont {M.}~\bibnamefont
  {Campanelli}}, \bibinfo {author} {\bibfnamefont {C.~O.}\ \bibnamefont
  {Lousto}}, \bibinfo {author} {\bibfnamefont {Y.}~\bibnamefont {Zlochower}}, \
  and\ \bibinfo {author} {\bibfnamefont {D.}~\bibnamefont {Merritt}},\ }\href
  {\doibase 10.1103/PhysRevLett.98.231102} {\bibfield  {journal} {\bibinfo
  {journal} {Phys. Rev. Lett.}\ }\textbf {\bibinfo {volume} {98}},\ \bibinfo
  {pages} {231102} (\bibinfo {year} {2007})},\ \Eprint
  {http://arxiv.org/abs/gr-qc/0702133} {arXiv:gr-qc/0702133} \BibitemShut
  {NoStop}%
\bibitem [{\citenamefont {Bruegmann}\ \emph {et~al.}(2008)\citenamefont
  {Bruegmann}, \citenamefont {Gonzalez}, \citenamefont {Hannam}, \citenamefont
  {Husa},\ and\ \citenamefont {Sperhake}}]{Brugmann:2007zj}%
  \BibitemOpen
  \bibfield  {author} {\bibinfo {author} {\bibfnamefont {B.}~\bibnamefont
  {Bruegmann}}, \bibinfo {author} {\bibfnamefont {J.~A.}\ \bibnamefont
  {Gonzalez}}, \bibinfo {author} {\bibfnamefont {M.}~\bibnamefont {Hannam}},
  \bibinfo {author} {\bibfnamefont {S.}~\bibnamefont {Husa}}, \ and\ \bibinfo
  {author} {\bibfnamefont {U.}~\bibnamefont {Sperhake}},\ }\href {\doibase
  10.1103/PhysRevD.77.124047} {\bibfield  {journal} {\bibinfo  {journal} {Phys.
  Rev. D}\ }\textbf {\bibinfo {volume} {77}},\ \bibinfo {pages} {124047}
  (\bibinfo {year} {2008})},\ \Eprint {http://arxiv.org/abs/0707.0135}
  {arXiv:0707.0135 [gr-qc]} \BibitemShut {NoStop}%
\bibitem [{\citenamefont {Healy}\ \emph {et~al.}(2009)\citenamefont {Healy},
  \citenamefont {Herrmann}, \citenamefont {Hinder}, \citenamefont {Shoemaker},
  \citenamefont {Laguna},\ and\ \citenamefont {Matzner}}]{Healy:2008js}%
  \BibitemOpen
  \bibfield  {author} {\bibinfo {author} {\bibfnamefont {J.}~\bibnamefont
  {Healy}}, \bibinfo {author} {\bibfnamefont {F.}~\bibnamefont {Herrmann}},
  \bibinfo {author} {\bibfnamefont {I.}~\bibnamefont {Hinder}}, \bibinfo
  {author} {\bibfnamefont {D.~M.}\ \bibnamefont {Shoemaker}}, \bibinfo {author}
  {\bibfnamefont {P.}~\bibnamefont {Laguna}}, \ and\ \bibinfo {author}
  {\bibfnamefont {R.~A.}\ \bibnamefont {Matzner}},\ }\href {\doibase
  10.1103/PhysRevLett.102.041101} {\bibfield  {journal} {\bibinfo  {journal}
  {Phys. Rev. Lett.}\ }\textbf {\bibinfo {volume} {102}},\ \bibinfo {pages}
  {041101} (\bibinfo {year} {2009})},\ \Eprint {http://arxiv.org/abs/0807.3292}
  {arXiv:0807.3292 [gr-qc]} \BibitemShut {NoStop}%
\bibitem [{\citenamefont {Lousto}\ and\ \citenamefont
  {Zlochower}(2011{\natexlab{a}})}]{Lousto:2010xk}%
  \BibitemOpen
  \bibfield  {author} {\bibinfo {author} {\bibfnamefont {C.~O.}\ \bibnamefont
  {Lousto}}\ and\ \bibinfo {author} {\bibfnamefont {Y.}~\bibnamefont
  {Zlochower}},\ }\href {\doibase 10.1103/PhysRevD.83.024003} {\bibfield
  {journal} {\bibinfo  {journal} {Phys. Rev. D}\ }\textbf {\bibinfo {volume}
  {83}},\ \bibinfo {pages} {024003} (\bibinfo {year} {2011}{\natexlab{a}})},\
  \Eprint {http://arxiv.org/abs/1011.0593} {arXiv:1011.0593 [gr-qc]}
  \BibitemShut {NoStop}%
\bibitem [{\citenamefont {Lousto}\ and\ \citenamefont
  {Zlochower}(2011{\natexlab{b}})}]{Lousto:2011kp}%
  \BibitemOpen
  \bibfield  {author} {\bibinfo {author} {\bibfnamefont {C.~O.}\ \bibnamefont
  {Lousto}}\ and\ \bibinfo {author} {\bibfnamefont {Y.}~\bibnamefont
  {Zlochower}},\ }\href {\doibase 10.1103/PhysRevLett.107.231102} {\bibfield
  {journal} {\bibinfo  {journal} {Phys. Rev. Lett.}\ }\textbf {\bibinfo
  {volume} {107}},\ \bibinfo {pages} {231102} (\bibinfo {year}
  {2011}{\natexlab{b}})},\ \Eprint {http://arxiv.org/abs/1108.2009}
  {arXiv:1108.2009 [gr-qc]} \BibitemShut {NoStop}%
\bibitem [{\citenamefont {Gerosa}\ and\ \citenamefont
  {Sesana}(2015)}]{Gerosa:2014gja}%
  \BibitemOpen
  \bibfield  {author} {\bibinfo {author} {\bibfnamefont {D.}~\bibnamefont
  {Gerosa}}\ and\ \bibinfo {author} {\bibfnamefont {A.}~\bibnamefont
  {Sesana}},\ }\href {\doibase 10.1093/mnras/stu2049} {\bibfield  {journal}
  {\bibinfo  {journal} {Mon. Not. Roy. Astron. Soc.}\ }\textbf {\bibinfo
  {volume} {446}},\ \bibinfo {pages} {38} (\bibinfo {year} {2015})},\ \Eprint
  {http://arxiv.org/abs/1405.2072} {arXiv:1405.2072 [astro-ph.GA]} \BibitemShut
  {NoStop}%
\bibitem [{\citenamefont {Gerosa}\ \emph {et~al.}(2018)\citenamefont {Gerosa},
  \citenamefont {Hebert},\ and\ \citenamefont {Stein}}]{Gerosa:2018qay}%
  \BibitemOpen
  \bibfield  {author} {\bibinfo {author} {\bibfnamefont {D.}~\bibnamefont
  {Gerosa}}, \bibinfo {author} {\bibfnamefont {F.}~\bibnamefont {Hebert}}, \
  and\ \bibinfo {author} {\bibfnamefont {L.~C.}\ \bibnamefont {Stein}},\ }\href
  {\doibase 10.1103/PhysRevD.97.104049} {\bibfield  {journal} {\bibinfo
  {journal} {Phys. Rev. D}\ }\textbf {\bibinfo {volume} {97}},\ \bibinfo
  {pages} {104049} (\bibinfo {year} {2018})},\ \Eprint
  {http://arxiv.org/abs/1802.04276} {arXiv:1802.04276 [gr-qc]} \BibitemShut
  {NoStop}%
\bibitem [{\citenamefont {Sperhake}\ \emph {et~al.}(2020)\citenamefont
  {Sperhake}, \citenamefont {Rosca-Mead}, \citenamefont {Gerosa},\ and\
  \citenamefont {Berti}}]{Sperhake:2019wwo}%
  \BibitemOpen
  \bibfield  {author} {\bibinfo {author} {\bibfnamefont {U.}~\bibnamefont
  {Sperhake}}, \bibinfo {author} {\bibfnamefont {R.}~\bibnamefont
  {Rosca-Mead}}, \bibinfo {author} {\bibfnamefont {D.}~\bibnamefont {Gerosa}},
  \ and\ \bibinfo {author} {\bibfnamefont {E.}~\bibnamefont {Berti}},\ }\href
  {\doibase 10.1103/PhysRevD.101.024044} {\bibfield  {journal} {\bibinfo
  {journal} {Phys. Rev. D}\ }\textbf {\bibinfo {volume} {101}},\ \bibinfo
  {pages} {024044} (\bibinfo {year} {2020})},\ \Eprint
  {http://arxiv.org/abs/1910.01598} {arXiv:1910.01598 [gr-qc]} \BibitemShut
  {NoStop}%
\bibitem [{\citenamefont {Sperhake}\ \emph {et~al.}(2011)\citenamefont
  {Sperhake}, \citenamefont {Berti}, \citenamefont {Cardoso}, \citenamefont
  {Pretorius},\ and\ \citenamefont {Yunes}}]{Sperhake:2010uv}%
  \BibitemOpen
  \bibfield  {author} {\bibinfo {author} {\bibfnamefont {U.}~\bibnamefont
  {Sperhake}}, \bibinfo {author} {\bibfnamefont {E.}~\bibnamefont {Berti}},
  \bibinfo {author} {\bibfnamefont {V.}~\bibnamefont {Cardoso}}, \bibinfo
  {author} {\bibfnamefont {F.}~\bibnamefont {Pretorius}}, \ and\ \bibinfo
  {author} {\bibfnamefont {N.}~\bibnamefont {Yunes}},\ }\href {\doibase
  10.1103/PhysRevD.83.024037} {\bibfield  {journal} {\bibinfo  {journal} {Phys.
  Rev. D}\ }\textbf {\bibinfo {volume} {83}},\ \bibinfo {pages} {024037}
  (\bibinfo {year} {2011})},\ \Eprint {http://arxiv.org/abs/1011.3281}
  {arXiv:1011.3281 [gr-qc]} \BibitemShut {NoStop}%
\bibitem [{\citenamefont {De~Luca}\ \emph
  {et~al.}(2020{\natexlab{a}})\citenamefont {De~Luca}, \citenamefont
  {Franciolini}, \citenamefont {Pani},\ and\ \citenamefont
  {Riotto}}]{DeLuca:2020qqa}%
  \BibitemOpen
  \bibfield  {author} {\bibinfo {author} {\bibfnamefont {V.}~\bibnamefont
  {De~Luca}}, \bibinfo {author} {\bibfnamefont {G.}~\bibnamefont
  {Franciolini}}, \bibinfo {author} {\bibfnamefont {P.}~\bibnamefont {Pani}}, \
  and\ \bibinfo {author} {\bibfnamefont {A.}~\bibnamefont {Riotto}},\
  }\href@noop {} {\  (\bibinfo {year} {2020}{\natexlab{a}})},\ \Eprint
  {http://arxiv.org/abs/2005.05641} {arXiv:2005.05641 [astro-ph.CO]}
  \BibitemShut {NoStop}%
\bibitem [{\citenamefont {Cardoso}\ \emph {et~al.}(2020)\citenamefont
  {Cardoso}, \citenamefont {Guo}, \citenamefont {Macedo},\ and\ \citenamefont
  {Pani}}]{Cardoso:2020nst}%
  \BibitemOpen
  \bibfield  {author} {\bibinfo {author} {\bibfnamefont {V.}~\bibnamefont
  {Cardoso}}, \bibinfo {author} {\bibfnamefont {W.-d.}\ \bibnamefont {Guo}},
  \bibinfo {author} {\bibfnamefont {C.~F.}\ \bibnamefont {Macedo}}, \ and\
  \bibinfo {author} {\bibfnamefont {P.}~\bibnamefont {Pani}},\ }\href@noop {}
  {\  (\bibinfo {year} {2020})},\ \Eprint {http://arxiv.org/abs/2009.07287}
  {arXiv:2009.07287 [gr-qc]} \BibitemShut {NoStop}%
\bibitem [{\citenamefont {Frolov}\ \emph {et~al.}(2018)\citenamefont {Frolov},
  \citenamefont {Krtouv~s},\ and\ \citenamefont {Kubizvnak}}]{Frolov:2018pys}%
  \BibitemOpen
  \bibfield  {author} {\bibinfo {author} {\bibfnamefont {V.~P.}\ \bibnamefont
  {Frolov}}, \bibinfo {author} {\bibfnamefont {P.}~\bibnamefont {Krtouv~s}}, \
  and\ \bibinfo {author} {\bibfnamefont {D.}~\bibnamefont {Kubizvnak}},\ }\href
  {\doibase 10.1103/PhysRevD.97.101701} {\bibfield  {journal} {\bibinfo
  {journal} {Phys. Rev. D}\ }\textbf {\bibinfo {volume} {97}},\ \bibinfo
  {pages} {101701} (\bibinfo {year} {2018})},\ \Eprint
  {http://arxiv.org/abs/1802.09491} {arXiv:1802.09491 [hep-th]} \BibitemShut
  {NoStop}%
\bibitem [{\citenamefont {Dolan}(2018)}]{Dolan:2018dqv}%
  \BibitemOpen
  \bibfield  {author} {\bibinfo {author} {\bibfnamefont {S.~R.}\ \bibnamefont
  {Dolan}},\ }\href {\doibase 10.1103/PhysRevD.98.104006} {\bibfield  {journal}
  {\bibinfo  {journal} {Phys. Rev. D}\ }\textbf {\bibinfo {volume} {98}},\
  \bibinfo {pages} {104006} (\bibinfo {year} {2018})},\ \Eprint
  {http://arxiv.org/abs/1806.01604} {arXiv:1806.01604 [gr-qc]} \BibitemShut
  {NoStop}%
\bibitem [{\citenamefont {Cayuso}\ \emph {et~al.}(2020)\citenamefont {Cayuso},
  \citenamefont {Dias}, \citenamefont {Gray}, \citenamefont {Kubizvnak},
  \citenamefont {Margalit}, \citenamefont {Santos}, \citenamefont
  {Gomes~Souza},\ and\ \citenamefont {Thiele}}]{Cayuso:2019ieu}%
  \BibitemOpen
  \bibfield  {author} {\bibinfo {author} {\bibfnamefont {R.}~\bibnamefont
  {Cayuso}}, \bibinfo {author} {\bibfnamefont {O.~J.}\ \bibnamefont {Dias}},
  \bibinfo {author} {\bibfnamefont {F.}~\bibnamefont {Gray}}, \bibinfo {author}
  {\bibfnamefont {D.}~\bibnamefont {Kubizvnak}}, \bibinfo {author}
  {\bibfnamefont {A.}~\bibnamefont {Margalit}}, \bibinfo {author}
  {\bibfnamefont {J.~E.}\ \bibnamefont {Santos}}, \bibinfo {author}
  {\bibfnamefont {R.}~\bibnamefont {Gomes~Souza}}, \ and\ \bibinfo {author}
  {\bibfnamefont {L.}~\bibnamefont {Thiele}},\ }\href {\doibase
  10.1007/JHEP04(2020)159} {\bibfield  {journal} {\bibinfo  {journal} {JHEP}\
  }\textbf {\bibinfo {volume} {04}},\ \bibinfo {pages} {159} (\bibinfo {year}
  {2020})},\ \Eprint {http://arxiv.org/abs/1912.08224} {arXiv:1912.08224
  [hep-th]} \BibitemShut {NoStop}%
\bibitem [{\citenamefont {Pani}\ \emph {et~al.}(2012)\citenamefont {Pani},
  \citenamefont {Cardoso}, \citenamefont {Gualtieri}, \citenamefont {Berti},\
  and\ \citenamefont {Ishibashi}}]{Pani:2012bp}%
  \BibitemOpen
  \bibfield  {author} {\bibinfo {author} {\bibfnamefont {P.}~\bibnamefont
  {Pani}}, \bibinfo {author} {\bibfnamefont {V.}~\bibnamefont {Cardoso}},
  \bibinfo {author} {\bibfnamefont {L.}~\bibnamefont {Gualtieri}}, \bibinfo
  {author} {\bibfnamefont {E.}~\bibnamefont {Berti}}, \ and\ \bibinfo {author}
  {\bibfnamefont {A.}~\bibnamefont {Ishibashi}},\ }\href {\doibase
  10.1103/PhysRevD.86.104017} {\bibfield  {journal} {\bibinfo  {journal} {Phys.
  Rev. D}\ }\textbf {\bibinfo {volume} {86}},\ \bibinfo {pages} {104017}
  (\bibinfo {year} {2012})},\ \Eprint {http://arxiv.org/abs/1209.0773}
  {arXiv:1209.0773 [gr-qc]} \BibitemShut {NoStop}%
\bibitem [{\citenamefont {Endlich}\ and\ \citenamefont
  {Penco}(2017)}]{Endlich:2016jgc}%
  \BibitemOpen
  \bibfield  {author} {\bibinfo {author} {\bibfnamefont {S.}~\bibnamefont
  {Endlich}}\ and\ \bibinfo {author} {\bibfnamefont {R.}~\bibnamefont
  {Penco}},\ }\href {\doibase 10.1007/JHEP05(2017)052} {\bibfield  {journal}
  {\bibinfo  {journal} {JHEP}\ }\textbf {\bibinfo {volume} {05}},\ \bibinfo
  {pages} {052} (\bibinfo {year} {2017})},\ \Eprint
  {http://arxiv.org/abs/1609.06723} {arXiv:1609.06723 [hep-th]} \BibitemShut
  {NoStop}%
\bibitem [{\citenamefont {Baryakhtar}\ \emph {et~al.}(2017)\citenamefont
  {Baryakhtar}, \citenamefont {Lasenby},\ and\ \citenamefont
  {Teo}}]{Baryakhtar:2017ngi}%
  \BibitemOpen
  \bibfield  {author} {\bibinfo {author} {\bibfnamefont {M.}~\bibnamefont
  {Baryakhtar}}, \bibinfo {author} {\bibfnamefont {R.}~\bibnamefont {Lasenby}},
  \ and\ \bibinfo {author} {\bibfnamefont {M.}~\bibnamefont {Teo}},\ }\href
  {\doibase 10.1103/PhysRevD.96.035019} {\bibfield  {journal} {\bibinfo
  {journal} {Phys. Rev. D}\ }\textbf {\bibinfo {volume} {96}},\ \bibinfo
  {pages} {035019} (\bibinfo {year} {2017})},\ \Eprint
  {http://arxiv.org/abs/1704.05081} {arXiv:1704.05081 [hep-ph]} \BibitemShut
  {NoStop}%
\bibitem [{\citenamefont {Baumann}\ \emph {et~al.}(2019)\citenamefont
  {Baumann}, \citenamefont {Chia}, \citenamefont {Stout},\ and\ \citenamefont
  {ter Haar}}]{Baumann:2019eav}%
  \BibitemOpen
  \bibfield  {author} {\bibinfo {author} {\bibfnamefont {D.}~\bibnamefont
  {Baumann}}, \bibinfo {author} {\bibfnamefont {H.~S.}\ \bibnamefont {Chia}},
  \bibinfo {author} {\bibfnamefont {J.}~\bibnamefont {Stout}}, \ and\ \bibinfo
  {author} {\bibfnamefont {L.}~\bibnamefont {ter Haar}},\ }\href {\doibase
  10.1088/1475-7516/2019/12/006} {\bibfield  {journal} {\bibinfo  {journal}
  {JCAP}\ }\textbf {\bibinfo {volume} {12}},\ \bibinfo {pages} {006} (\bibinfo
  {year} {2019})},\ \Eprint {http://arxiv.org/abs/1908.10370} {arXiv:1908.10370
  [gr-qc]} \BibitemShut {NoStop}%
\bibitem [{\citenamefont {East}(2017)}]{East:2017mrj}%
  \BibitemOpen
  \bibfield  {author} {\bibinfo {author} {\bibfnamefont {W.~E.}\ \bibnamefont
  {East}},\ }\href {\doibase 10.1103/PhysRevD.96.024004} {\bibfield  {journal}
  {\bibinfo  {journal} {Phys. Rev. D}\ }\textbf {\bibinfo {volume} {96}},\
  \bibinfo {pages} {024004} (\bibinfo {year} {2017})},\ \Eprint
  {http://arxiv.org/abs/1705.01544} {arXiv:1705.01544 [gr-qc]} \BibitemShut
  {NoStop}%
\bibitem [{\citenamefont {Cardoso}\ \emph {et~al.}(2018)\citenamefont
  {Cardoso}, \citenamefont {Dias}, \citenamefont {Hartnett}, \citenamefont
  {Middleton}, \citenamefont {Pani},\ and\ \citenamefont
  {Santos}}]{Cardoso:2018tly}%
  \BibitemOpen
  \bibfield  {author} {\bibinfo {author} {\bibfnamefont {V.}~\bibnamefont
  {Cardoso}}, \bibinfo {author} {\bibfnamefont {O.~J.}\ \bibnamefont {Dias}},
  \bibinfo {author} {\bibfnamefont {G.~S.}\ \bibnamefont {Hartnett}}, \bibinfo
  {author} {\bibfnamefont {M.}~\bibnamefont {Middleton}}, \bibinfo {author}
  {\bibfnamefont {P.}~\bibnamefont {Pani}}, \ and\ \bibinfo {author}
  {\bibfnamefont {J.~E.}\ \bibnamefont {Santos}},\ }\href {\doibase
  10.1088/1475-7516/2018/03/043} {\bibfield  {journal} {\bibinfo  {journal}
  {JCAP}\ }\textbf {\bibinfo {volume} {03}},\ \bibinfo {pages} {043} (\bibinfo
  {year} {2018})},\ \Eprint {http://arxiv.org/abs/1801.01420} {arXiv:1801.01420
  [gr-qc]} \BibitemShut {NoStop}%
\bibitem [{\citenamefont {Yoshino}\ and\ \citenamefont
  {Kodama}(2012)}]{Yoshino:2012kn}%
  \BibitemOpen
  \bibfield  {author} {\bibinfo {author} {\bibfnamefont {H.}~\bibnamefont
  {Yoshino}}\ and\ \bibinfo {author} {\bibfnamefont {H.}~\bibnamefont
  {Kodama}},\ }\href {\doibase 10.1143/PTP.128.153} {\bibfield  {journal}
  {\bibinfo  {journal} {Prog. Theor. Phys.}\ }\textbf {\bibinfo {volume}
  {128}},\ \bibinfo {pages} {153} (\bibinfo {year} {2012})},\ \Eprint
  {http://arxiv.org/abs/1203.5070} {arXiv:1203.5070 [gr-qc]} \BibitemShut
  {NoStop}%
\bibitem [{\citenamefont {Ikeda}\ \emph {et~al.}(2019)\citenamefont {Ikeda},
  \citenamefont {Brito},\ and\ \citenamefont {Cardoso}}]{Ikeda:2018nhb}%
  \BibitemOpen
  \bibfield  {author} {\bibinfo {author} {\bibfnamefont {T.}~\bibnamefont
  {Ikeda}}, \bibinfo {author} {\bibfnamefont {R.}~\bibnamefont {Brito}}, \ and\
  \bibinfo {author} {\bibfnamefont {V.}~\bibnamefont {Cardoso}},\ }\href
  {\doibase 10.1103/PhysRevLett.122.081101} {\bibfield  {journal} {\bibinfo
  {journal} {Phys. Rev. Lett.}\ }\textbf {\bibinfo {volume} {122}},\ \bibinfo
  {pages} {081101} (\bibinfo {year} {2019})},\ \Eprint
  {http://arxiv.org/abs/1811.04950} {arXiv:1811.04950 [gr-qc]} \BibitemShut
  {NoStop}%
\bibitem [{\citenamefont {Blas}\ and\ \citenamefont
  {Witte}(2020)}]{BlasWitte_axionSRs}%
  \BibitemOpen
  \bibfield  {author} {\bibinfo {author} {\bibfnamefont {D.}~\bibnamefont
  {Blas}}\ and\ \bibinfo {author} {\bibfnamefont {S.~J.}\ \bibnamefont
  {Witte}},\ }\href {\doibase 10.1103/PhysRevD.102.103018} {\bibfield
  {journal} {\bibinfo  {journal} {Phys. Rev. D}\ }\textbf {\bibinfo {volume}
  {102}},\ \bibinfo {pages} {103018} (\bibinfo {year} {2020})},\ \Eprint
  {http://arxiv.org/abs/2009.10074} {arXiv:2009.10074 [astro-ph.CO]}
  \BibitemShut {NoStop}%
\bibitem [{\citenamefont {Mathur}\ \emph {et~al.}(2020)\citenamefont {Mathur},
  \citenamefont {Rajendran},\ and\ \citenamefont {Tanin}}]{Mathur:2020aqv}%
  \BibitemOpen
  \bibfield  {author} {\bibinfo {author} {\bibfnamefont {A.}~\bibnamefont
  {Mathur}}, \bibinfo {author} {\bibfnamefont {S.}~\bibnamefont {Rajendran}}, \
  and\ \bibinfo {author} {\bibfnamefont {E.~H.}\ \bibnamefont {Tanin}},\
  }\href@noop {} {\  (\bibinfo {year} {2020})},\ \Eprint
  {http://arxiv.org/abs/2004.12326} {arXiv:2004.12326 [hep-ph]} \BibitemShut
  {NoStop}%
\bibitem [{\citenamefont {Schwinger}(1951)}]{schwinger1951gauge}%
  \BibitemOpen
  \bibfield  {author} {\bibinfo {author} {\bibfnamefont {J.}~\bibnamefont
  {Schwinger}},\ }\href@noop {} {\bibfield  {journal} {\bibinfo  {journal}
  {Physical Review}\ }\textbf {\bibinfo {volume} {82}},\ \bibinfo {pages} {664}
  (\bibinfo {year} {1951})}\BibitemShut {NoStop}%
\bibitem [{\citenamefont {Raffelt}(1996)}]{raffelt1996stars}%
  \BibitemOpen
  \bibfield  {author} {\bibinfo {author} {\bibfnamefont {G.~G.}\ \bibnamefont
  {Raffelt}},\ }\href@noop {} {\emph {\bibinfo {title} {Stars as laboratories
  for fundamental physics: The astrophysics of neutrinos, axions, and other
  weakly interacting particles}}}\ (\bibinfo  {publisher} {University of
  Chicago press},\ \bibinfo {year} {1996})\BibitemShut {NoStop}%
\bibitem [{\citenamefont {Gnedin}\ and\ \citenamefont
  {Hui}(1997)}]{gnedin1997probing}%
  \BibitemOpen
  \bibfield  {author} {\bibinfo {author} {\bibfnamefont {N.~Y.}\ \bibnamefont
  {Gnedin}}\ and\ \bibinfo {author} {\bibfnamefont {L.}~\bibnamefont {Hui}},\
  }\href@noop {} {\bibfield  {journal} {\bibinfo  {journal} {arXiv preprint
  astro-ph/9706219}\ } (\bibinfo {year} {1997})}\BibitemShut {NoStop}%
\bibitem [{\citenamefont {Bertschinger}(1998)}]{bertschinger1998simulations}%
  \BibitemOpen
  \bibfield  {author} {\bibinfo {author} {\bibfnamefont {E.}~\bibnamefont
  {Bertschinger}},\ }\href@noop {} {\bibfield  {journal} {\bibinfo  {journal}
  {Annual Review of Astronomy and Astrophysics}\ }\textbf {\bibinfo {volume}
  {36}},\ \bibinfo {pages} {599} (\bibinfo {year} {1998})}\BibitemShut
  {NoStop}%
\bibitem [{\citenamefont {Witte}\ \emph {et~al.}(2020)\citenamefont {Witte},
  \citenamefont {Rosauro-Alcaraz}, \citenamefont {McDermott},\ and\
  \citenamefont {Poulin}}]{Witte:2020rvb}%
  \BibitemOpen
  \bibfield  {author} {\bibinfo {author} {\bibfnamefont {S.~J.}\ \bibnamefont
  {Witte}}, \bibinfo {author} {\bibfnamefont {S.}~\bibnamefont
  {Rosauro-Alcaraz}}, \bibinfo {author} {\bibfnamefont {S.~D.}\ \bibnamefont
  {McDermott}}, \ and\ \bibinfo {author} {\bibfnamefont {V.}~\bibnamefont
  {Poulin}},\ }\href {\doibase 10.1007/JHEP06(2020)132} {\bibfield  {journal}
  {\bibinfo  {journal} {JHEP}\ }\textbf {\bibinfo {volume} {06}},\ \bibinfo
  {pages} {132} (\bibinfo {year} {2020})},\ \Eprint
  {http://arxiv.org/abs/2003.13698} {arXiv:2003.13698 [astro-ph.CO]}
  \BibitemShut {NoStop}%
\bibitem [{\citenamefont {Braaten}\ and\ \citenamefont
  {Segel}(1993)}]{braaten1993neutrino}%
  \BibitemOpen
  \bibfield  {author} {\bibinfo {author} {\bibfnamefont {E.}~\bibnamefont
  {Braaten}}\ and\ \bibinfo {author} {\bibfnamefont {D.}~\bibnamefont
  {Segel}},\ }\href@noop {} {\bibfield  {journal} {\bibinfo  {journal}
  {Physical Review D}\ }\textbf {\bibinfo {volume} {48}},\ \bibinfo {pages}
  {1478} (\bibinfo {year} {1993})}\BibitemShut {NoStop}%
\bibitem [{\citenamefont {Kaw}\ and\ \citenamefont
  {Dawson}(1970)}]{kaw1970relativistic}%
  \BibitemOpen
  \bibfield  {author} {\bibinfo {author} {\bibfnamefont {P.}~\bibnamefont
  {Kaw}}\ and\ \bibinfo {author} {\bibfnamefont {J.}~\bibnamefont {Dawson}},\
  }\href@noop {} {\bibfield  {journal} {\bibinfo  {journal} {The Physics of
  Fluids}\ }\textbf {\bibinfo {volume} {13}},\ \bibinfo {pages} {472} (\bibinfo
  {year} {1970})}\BibitemShut {NoStop}%
\bibitem [{\citenamefont {Max}\ and\ \citenamefont
  {Perkins}(1971)}]{max1971strong}%
  \BibitemOpen
  \bibfield  {author} {\bibinfo {author} {\bibfnamefont {C.}~\bibnamefont
  {Max}}\ and\ \bibinfo {author} {\bibfnamefont {F.}~\bibnamefont {Perkins}},\
  }\href@noop {} {\bibfield  {journal} {\bibinfo  {journal} {Physical Review
  Letters}\ }\textbf {\bibinfo {volume} {27}},\ \bibinfo {pages} {1342}
  (\bibinfo {year} {1971})}\BibitemShut {NoStop}%
\bibitem [{\citenamefont {Sauter}(1931)}]{sauter1931verhalten}%
  \BibitemOpen
  \bibfield  {author} {\bibinfo {author} {\bibfnamefont {F.}~\bibnamefont
  {Sauter}},\ }\href@noop {} {\bibfield  {journal} {\bibinfo  {journal}
  {Zeitschrift f{\"u}r Physik}\ }\textbf {\bibinfo {volume} {69}},\ \bibinfo
  {pages} {742} (\bibinfo {year} {1931})}\BibitemShut {NoStop}%
\bibitem [{\citenamefont {Euler}\ and\ \citenamefont
  {Heisenberg}(1936)}]{euler1936consequences}%
  \BibitemOpen
  \bibfield  {author} {\bibinfo {author} {\bibfnamefont {H.}~\bibnamefont
  {Euler}}\ and\ \bibinfo {author} {\bibfnamefont {W.}~\bibnamefont
  {Heisenberg}},\ }\href@noop {} {\bibfield  {journal} {\bibinfo  {journal} {Z.
  Phys}\ }\textbf {\bibinfo {volume} {98}},\ \bibinfo {pages} {714} (\bibinfo
  {year} {1936})}\BibitemShut {NoStop}%
\bibitem [{\citenamefont {Heisenberg}\ and\ \citenamefont
  {Euler}(2006)}]{heisenberg2006consequences}%
  \BibitemOpen
  \bibfield  {author} {\bibinfo {author} {\bibfnamefont {W.}~\bibnamefont
  {Heisenberg}}\ and\ \bibinfo {author} {\bibfnamefont {H.}~\bibnamefont
  {Euler}},\ }\href@noop {} {\bibfield  {journal} {\bibinfo  {journal} {arXiv
  preprint physics/0605038}\ } (\bibinfo {year} {2006})}\BibitemShut {NoStop}%
\bibitem [{\citenamefont {De~Luca}\ \emph
  {et~al.}(2020{\natexlab{b}})\citenamefont {De~Luca}, \citenamefont
  {Franciolini}, \citenamefont {Pani},\ and\ \citenamefont
  {Riotto}}]{DeLuca:2020bjf}%
  \BibitemOpen
  \bibfield  {author} {\bibinfo {author} {\bibfnamefont {V.}~\bibnamefont
  {De~Luca}}, \bibinfo {author} {\bibfnamefont {G.}~\bibnamefont
  {Franciolini}}, \bibinfo {author} {\bibfnamefont {P.}~\bibnamefont {Pani}}, \
  and\ \bibinfo {author} {\bibfnamefont {A.}~\bibnamefont {Riotto}},\ }\href
  {\doibase 10.1088/1475-7516/2020/04/052} {\bibfield  {journal} {\bibinfo
  {journal} {JCAP}\ }\textbf {\bibinfo {volume} {04}},\ \bibinfo {pages} {052}
  (\bibinfo {year} {2020}{\natexlab{b}})},\ \Eprint
  {http://arxiv.org/abs/2003.02778} {arXiv:2003.02778 [astro-ph.CO]}
  \BibitemShut {NoStop}%
\bibitem [{\citenamefont {Draine}(2010)}]{draine2010physics}%
  \BibitemOpen
  \bibfield  {author} {\bibinfo {author} {\bibfnamefont {B.~T.}\ \bibnamefont
  {Draine}},\ }\href@noop {} {\emph {\bibinfo {title} {Physics of the
  interstellar and intergalactic medium}}}\ (\bibinfo  {publisher} {Princeton
  University Press},\ \bibinfo {year} {2010})\BibitemShut {NoStop}%
\bibitem [{\citenamefont {McDermott}\ and\ \citenamefont
  {Witte}(2020)}]{McDermott:2019lch}%
  \BibitemOpen
  \bibfield  {author} {\bibinfo {author} {\bibfnamefont {S.~D.}\ \bibnamefont
  {McDermott}}\ and\ \bibinfo {author} {\bibfnamefont {S.~J.}\ \bibnamefont
  {Witte}},\ }\href {\doibase 10.1103/PhysRevD.101.063030} {\bibfield
  {journal} {\bibinfo  {journal} {Phys. Rev. D}\ }\textbf {\bibinfo {volume}
  {101}},\ \bibinfo {pages} {063030} (\bibinfo {year} {2020})},\ \Eprint
  {http://arxiv.org/abs/1911.05086} {arXiv:1911.05086 [hep-ph]} \BibitemShut
  {NoStop}%
\bibitem [{\citenamefont {Mo}\ \emph {et~al.}(2010)\citenamefont {Mo},
  \citenamefont {Van~den Bosch},\ and\ \citenamefont {White}}]{mo2010galaxy}%
  \BibitemOpen
  \bibfield  {author} {\bibinfo {author} {\bibfnamefont {H.}~\bibnamefont
  {Mo}}, \bibinfo {author} {\bibfnamefont {F.}~\bibnamefont {Van~den Bosch}}, \
  and\ \bibinfo {author} {\bibfnamefont {S.}~\bibnamefont {White}},\
  }\href@noop {} {\emph {\bibinfo {title} {Galaxy formation and evolution}}}\
  (\bibinfo  {publisher} {Cambridge University Press},\ \bibinfo {year}
  {2010})\BibitemShut {NoStop}%
\bibitem [{\citenamefont {Mendoza}(1983)}]{mendoza1983planetary}%
  \BibitemOpen
  \bibfield  {author} {\bibinfo {author} {\bibfnamefont {C.}~\bibnamefont
  {Mendoza}},\ }in\ \href@noop {} {\emph {\bibinfo {booktitle} {IAU Symp}}},\
  Vol.\ \bibinfo {volume} {103}\ (\bibinfo {year} {1983})\ p.\ \bibinfo {pages}
  {143}\BibitemShut {NoStop}%
\bibitem [{\citenamefont {Longair}(2011)}]{longair2011high}%
  \BibitemOpen
  \bibfield  {author} {\bibinfo {author} {\bibfnamefont {M.~S.}\ \bibnamefont
  {Longair}},\ }\href@noop {} {\emph {\bibinfo {title} {High energy
  astrophysics}}}\ (\bibinfo  {publisher} {Cambridge university press},\
  \bibinfo {year} {2011})\BibitemShut {NoStop}%
\end{thebibliography}%
%%%%%%%%%%%%%%%%%%%%%%%%%%%%%%%%%%%%%%%%%%%%%%%%%%%%%

\end{document}